\documentclass{aa}

\usepackage{natbib}
\bibpunct{(}{)}{;}{a}{}{,} 

\usepackage{array, multirow, graphicx}
\usepackage[dvipsnames]{xcolor}
\usepackage{wasysym}[integrals]
\usepackage{amssymb}
\usepackage{placeins}
\usepackage{pifont}
\usepackage{siunitx}
\usepackage{booktabs, tabularx}
\usepackage{cellspace, makecell} 
\usepackage{mathtools}
\usepackage{txfonts}

\usepackage{csquotes}

\usepackage[]{hyperref}
\hypersetup{colorlinks,citecolor=blue, linkcolor=blue, urlcolor=blue, breaklinks=true}

\newcounter{magicrownumbers}

\begin{document} 
%%%%%%%%%%%%%%%%%%%%%%%%%%%%%%%%%%%%%%%%%
%%%%%%%%%%NUOVI COMANDI%%%%%%%%%%%%%%%%%%%%%%
\newcommand{\bcdot}{\boldsymbol{\cdot}}%prodotto scalare
\newcommand{\w}[1]{\mathbf{#1}}%vettore grassetto
\newcommand{\bra}[1]{\left\langle #1 \middle | \right.}%notazione Dirac
\newcommand{\ket}[1]{\left.\middle | #1\right\rangle}%notazione Dirac
\newcommand{\braket}[2]{\left\langle #1  \middle | #2 \right\rangle}%prodotto Dirac
\newcommand{\oplin}[2]{\left.\middle | #1 \right \rangle \left \langle #2 \middle | \right.}
\newcommand{\lf}{\left}%sinistra
\newcommand{\rg}{\right}%destra
\newcommand{\tonda}[1]{\!\left ( #1 \right )}%parentesi tonda
\newcommand{\quadra}[1]{\left [ #1 \right ]}%parentesi quadra
\newcommand{\graffa}[1]{\left \{ #1 \right \}}%parentesi graffa
\newcommand{\dirac}[1]{\updelta\!\left( #1 \right )}%delta di Dirac
\newcommand{\deltak}[1]{\updelta_{ #1 }}%delta di Kronecher
\newcommand{\D}[2]{\frac{d #1}{d  #2}}%derivata prima
\newcommand{\DD}[2]{\frac{d^2 #1}{d #2^2}}%derivata seconda
\newcommand{\pd}[2]{\frac{\partial #1}{\partial #2}}%derivata parziale prima
\newcommand{\pdd}[2]{\frac{\partial^2 #1}{\partial #2^2}}%derivata parziale seconda
\newcommand{\medio}[1]{\left\langle #1 \right\rangle}%valore medio
\newcommand{\abs}[1]{\left | #1 \right |}%valore assoluto
\newcommand{\norma}[1]{\left \| #1 \right \|}%norma
\newcommand{\sca}[2]{\left \langle #1 , #2 \right\rangle}%prodotto scalare 2
\newcommand{\ndiv}[1]{{\boldsymbol{\nabla}}\boldsymbol{\cdot}\w{#1}}%divergenza
\newcommand{\ndivx}[1]{{\boldsymbol{\nabla}_{\w{x}}}\boldsymbol{\cdot}\w{#1}}
\newcommand{\ndivv}[1]{{\boldsymbol{\nabla}_{\w{v}}}\boldsymbol{\cdot}\w{#1}}
\newcommand{\grad}{{\boldsymbol{\nabla}}}%gradiente
\newcommand{\gradx}{{\boldsymbol{\nabla}}_{\w{x}}}
\newcommand{\gradv}{{\boldsymbol{\nabla}}_{\w{v}}}
\newcommand{\lap}[1]{\boldsymbol{\nabla}^2 #1}%laplaciano
\newcommand{\rot}[1]{\boldsymbol{\nabla}\times\w{#1}}%rotore
\newcommand{\pois}[2]{\Bigl \{ #1 , #2 \Bigr\}}%parentesi di poisson
\newcommand{\com}[2]{\left [ #1 , #2 \right ]}%commutatore
\newcommand{\sistemai}{\lf\{\begin{aligned}}
\newcommand{\sistemaf}{\end{aligned}\rg.}
\newcommand{\implica}{\Longrightarrow}%% freccia grossa di "implica"
\newcommand{\sse}{\Longleftrightarrow}%freccia se e solo se
\newcommand{\id}{\mathbb{I}}%matrice identit
\newcommand{\sopra}[1]{\overline{#1}}
\newcommand{\sotto}[1]{\underline{#1}}
\newcommand{\gv}[1]{\ensuremath{\mbox{\boldmath$ #1 $}}}
\newcommand{\calc}[3]{\lf.{#1}\rg |_{#2}^{#3}}%calcolato in..
\newcommand{\cro}{\dagger}%croce
\newcommand{\eq}[1]{\begin{equation} #1 \end{equation}}
\newcommand{\eqn}[1]{\begin{equation*} #1 \end{equation*}}
\newcommand{\virg}[1]{``#1''}
\newcommand{\dmat}[1]{\frac{\mathcal{D} #1}{\mathcal{D}t}}
\newcommand{\mc}[1]{\mathcal{#1}}
\newcommand{\bxi}{\boldsymbol{\xi}}
\newcommand{\vx}{\hat{\w{x}}}
\newcommand{\vy}{\hat{\w{y}}}
\newcommand{\vz}{\hat{\w{z}}}
\newcommand{\vn}{\hat{\w{n}}}
\newcommand{\disc}[1]{\biggl[ #1 \biggr]}
\newcommand{\td}{\tau_{\text{diff}}}
\newcommand{\tc}{\tau_{\text{conv}}}
\newcommand{\ft}{\tilde{f}_1}
\newcommand{\fth}{\widehat{f}_1}
\newcommand{\et}{\tilde{\w{E}}_1}
\newcommand{\eet}{\tilde{E}_1}
\newcommand{\eeth}{\widehat{E}_1}
\newcommand{\asun}{\astrosun}
\newcommand{\al}[1]{\begin{aligned} #1 \end{aligned}}
\newcommand{\ftonda}[2]{\!\left ( \frac{#1}{#2} \right )}
\newcommand{\mum}{\unit{\mu m}}
\newcommand{\hh}{H$_{2}$ }
\newcommand{\Mbh}{M_\text{BH}}
\newcommand{\bs}[1]{\boldsymbol{#1}}
\newcommand{\fit}{\emph{fit}\xspace}
\newcommand{\pixel}{\emph{pixel}\xspace}
%% FSLs
\newcommand{\RNum}[1]{\uppercase\expandafter{\romannumeral #1\relax}}
\newcommand{\oi}{[O\RNum{1}]$_{\rm 63\,\mu m}$}
\newcommand{\oiii}[1]{[O\RNum{3}]$_{\rm #1\,\mu m}$}
\newcommand{\nii}[1]{[N\RNum{2}]$_{\rm #1\,\mu m}$}
\newcommand{\cii}{[C\RNum{2}]$_{\rm 158\,\mu m}$}
\newcommand{\ci}[1]{[C\RNum{1}]$_{\rm #1\,\mu m}$}
%% image folder
%\newcommand{\imagefolder}{/Users/antoniopensabene/MQN01/cube_cleaned_2.0s_multiscale.40kms.image.fits_prop}
%\newcommand{\mainfolder}{/Users/antoniopensabene/MQN01/}
\newcommand{\tabledir}{/Users/antoniopensabene/MEGA/Papers/MQN01-I/tables}
\newcommand{\imagesdir}{/Users/antoniopensabene/MEGA/Papers/MQN01-I/images}
\newcommand{\mosbthree}{/Users/antoniopensabene/MQN01_mosaicB3}
\newcommand{\mosbsix}{/Users/antoniopensabene/MQN01_mosaicB6}

   \title{Detection of metal absorption lines in quasar spectra}

   \subtitle{A neural network approach using U-Net}

   \author{Elena~Sofia~Mangola
          \inst{\ref{marseille}, \ref{unimib}}\thanks{Corresponding author: elena.mangola@lam.fr}
          \and
          Francesco~Pistis\inst{\ref{unimib}, \ref{ncbj}, \ref{inaf_bologna}}
          \and
          Michele~Fumagalli\inst{\ref{unimib}, \ref{inaf_trieste}}
          \and
          Matteo~Fossati\inst{\ref{unimib}, \ref{inaf_brera}}
          \and
          Ting-Yun Cheng\inst{\ref{kapteyn}}
          \and
          Ryan J. Cooke\inst{\ref{durham}}
          \and
          Rajeshwari Dutta\inst{\ref{iucaa}}
          \and
          Ignasi Pérez-Ràfols\inst{\ref{UPC}}
          \and
          Matthew Pieri\inst{\ref{marseille}}
          \and
          Emanuel Gafton\inst{\ref{ING}}
          }
          
\institute{
    Aix Marseille Univ. CNRS, CNES, LAM, Marseille, France\label{marseille}
    \and 
    Dipartimento di Fisica \enquote*{G. Occhialini}, Universit\`{a} degli Studi di Milano-Bicocca, Piazza della Scienza 3, 20126 Milano, Italy\label{unimib}
    \and
    National Centre for Nuclear Research, ul. Pasteura 7, 02-093 Warsaw, Poland\label{ncbj}
    \and
    INAF - Osservatorio di Astrofisica e Scienza dello Spazio di Bologna, Via Piero Gobetti 93/3, I-40129 Bologna, Italy\label{inaf_bologna}
    \and
    INAF - Osservatorio Astronomico di Trieste, via G.B. Tiepolo 11, I-34143 Trieste, Italy\label{inaf_trieste}
    \and
    INAF - Osservatorio Astronomico di Brera, Via Brera 28, 20122 Milano, via E. Bianchi 46, 23807 Merate, Italy\label{inaf_brera}
    \and
    Kapteyn Astronomical Institute, University of Groningen, Landleven 12 (Kapteynborg, 5419), 9747 AD Groningen, The Netherlands\label{kapteyn}
    \and
    Centre for Extragalactic Astronomy, Durham University, South Road, Durham DH1 3LE, UK\label{durham}
    \and
    IUCAA, Postbag 4, Ganeshkind, Pune 411007, India\label{iucaa}
    \and
    Departament de Física, EEBE, Universitat Politècnica de Catalunya, c/Eduard Maristany 10, 08930 Barcelona, Spain\label{UPC}
    \and
    Isaac Newton Group of Telescopes, Apartado 321, 38700 Santa Cruz de la Palma, Tenerife, Spain\label{ING}
}
   
\date{\today}

\abstract 
   {Current and future large spectroscopic surveys are significantly enhancing the volume and resolution of quasar spectra that are observed, which requires the creation of efficient and precise automated techniques to detect absorption features.}
   {This study focuses on the detection of metal absorption features using a novel U-Net model on WEAVE-like mock spectra in the quasar rest-frame wavelength interval $1\,230 \ \mathrm{\AA} \leq \lambda_\mathrm{RF} \leq 3\,095 \mathrm{\AA}$. We test the network performance for absorption detection both on ideal data and after simulating the continuum fitting step as applied on real data.}
   {The performance of these architectures is evaluated by the completeness, purity, and F1 score reached in bins of signal-to-noise ($\mathrm{S}/ \mathrm{N}_\mathrm{line} $) for the absorption lines and with the absolute fractional flux error for the continuum. The ability to recover the correct line centers is also studied.}
   {The U-Net reaches scores of $\approx 90\%$ for all metrics (completeness, purity, and F1 score) at $\mathrm{S}/ \mathrm{N}_\mathrm{line} \approx 4$. All false positive detections with $\mathrm{S}/ \mathrm{N}_\mathrm{line} \geq 5$ fall in the tails of the broad Ly$\alpha$ absorbers distribution of damped Ly$\alpha$ systems. The combination of continuum fitting and line detections has negligible effects on the detection performance at $\mathrm{S}/ \mathrm{N}_\mathrm{line} \geq 4$.}
   {Our proposed U-Net architecture offers a competitive tool for the analysis of absorption lines in current and upcoming large spectroscopic surveys and is well-suited to the identification of any absorption line feature.}


\maketitle

\section{Introduction}\label{sec:intro}

Quasar spectra enable the study of the intergalactic medium (IGM) through the footprints left in the form of absorption lines by hydrogen and metals.
For example, the neutral hydrogen distribution within the IGM can be observed at high redshift ($z\gtrsim1.5$) as a dense forest of absorber systems at wavelengths lower than $1\,216\ \mathrm{\AA}$ in the quasar rest-frame, the so-called Lyman-$\alpha$ forest \citep[Ly$\alpha$,][]{lynds1971abs, bahcall1971abs, mcdonald2006lyalpha, shull2012baryon}.
Quasar spectroscopy further enables the study of the composition and kinematics of the diffuse, often metal-enriched, circumgalactic medium (CGM), through the measurement of (partially) optically-thick hydrogen lines \citep[e.g,][]{2002ApJ...566...68P,2005ApJ...635..123P,2011MNRAS.416.1215R,2013ApJ...770..138L,2024MNRAS.532...32M} or strong ions such as \ion{C}{iv}, \ion{Mg}{II}, or \ion{Si}{iv}
\citep[e.g.,][]{hasan2020civ,Cooksey_2013, 2013ApJ...770..130Z, d'odorico2016metals, d'odorico2022siiv, gontcho2018civ, dutta2020magg, galbiati2023magg, galbiati2024magg, yu2025civ}.

The growing volume of data from large spectroscopic surveys, such as the Sloan Digital Sky Survey \citep[SDSS,][]{abazajian2009sdss}, the VIMOS Public Extragalactic Redshift Survey \citep[VIPERS,][]{scodeggio2018vimos}, the Dark Energy Spectroscopic Instrument \citep[DESI,][]{desi2022overview}, and the European Space Agency's Euclid mission \citep{laureijs2011euclid}, along with upcoming projects like the WHT Enhanced Area Velocity Explorer \citep[WEAVE,][]{dalton2012weave, dalton2014weave, dalton2016weave, jin2024weave} and the 4-meter Multi-Object Spectroscopic Telescope \citep[4MOST,][]{dejong20194most}, has made the task of detecting the absorption lines challenging for traditional detection techniques, such as kernel filtering \citep[e.g.,][]{anand2021qsoabs, zou2021qsoabs, zhu_menard}, and spectral template-fitting methods \citep[e.g.,][]{2025PhRvD.112h3510B}. Even though these approaches can provide accurate detections, their application to large spectroscopic catalogs can be limited due to the computational cost of applying filters to large datasets.
Therefore, it is essential to turn to fully automated algorithms that can deliver accurate and precise detection in extensive catalogs while using reduced computational resources and time.

Machine learning (ML) techniques have become ideal tools for fast and accurate detection of absorption lines in quasar spectra, as demonstrated by an increasing number of applications that can be found in the literature. Examples of the ML techniques used include Gaussian processes for the identification of damped Ly$\alpha$ systems \citep[DLAs,][]{Garnett_dla} and \ion{C}{iv} absorption lines detection \citep{2023MNRAS.526.4557M}, random forest classifiers for searches of Lyman limit systems \citep[LLSs,][]{2020MNRAS.498.1951F}, and artificial neural networks (ANN or NN) for the identification of DLAs \citep{2018MNRAS.476.1151P,2022ApJS..259...28W}, Ly$\alpha$ forest lines \citep{2022MNRAS.517..755C}, and \ion{Mg}{ii} absorption lines \citep[]{zhao_mgii,Szakacs_mgii}.

In this work, we build a new NN model aimed at detecting metal absorption lines in quasar spectra. For this task, we select a U-Net architecture \citep{ronneberger2015unet} for its ability to track both local and global structures in the data.
U-Net models have the ability to produce accurate semantic segmentations of images. They have been applied to several astrophysics problems like denoising \citep{vojtekova_2021}, reconstructing spectra and density fields \citep{zhong_2025, aragon_2019, 2025ApJS..280...53S}, and objects detection \citep{2024MNRAS.533.1426N, 2025ApJ...994..117S}.

The model is trained and tested using mock spectra for the WEAVE-QSO survey. The same mock data were used in the companion paper by \citet{pistis2025continuum} to derive a model for the quasar continuum instead.
WEAVE \citep{jin2024weave, Rogers2014} is a multiobject wide-field spectroscopic survey at the $4.2$m William Herschel Telescope at the Roque de los Muchachos Observatory in La Palma, Spain \citep{dalton2016weave, 2020SPIE11447E..14D}.
The WEAVE spectrograph is fed by $\sim1000$ fibers positioned within a 2-degree field of view, enabled by a prime focus corrector, and covers the $366-969\;\rm{nm}$ range at resolutions $R\sim5000$ and $20\,000$.
The WEAVE survey comprises eight individual surveys that address a variety of scientific goals, ranging from stellar to extragalactic astronomy.
In particular, the WEAVE-QSO survey \citep[WQ;][]{pieri2016weaveqso} is of interest for this work.
The WEAVE-QSO survey aims at observing $\gtrsim 10^5$ high-redshift ($z \geq 2.2$) quasars over an area of $\approx 6\,000\rm~deg^2$ \citep[termed WQ-Wide area,][]{jin2024weave}. In this work, we will refer primarily to this survey, but our results can be easily generalized to other surveys. 

The paper is organized as follows.
In Sect.~\ref{sec:data}, we briefly describe the WEAVE mock data used in this work, in Sect.~\ref{sec:method} we outline the data preprocessing and the construction of the NN model, while in Sect.~\ref{sec:application}, we detail the application of the model to the mock spectra. In Sect.~\ref{sec:comparison}, we discuss and compare our work with prior literature.
Finally, we discuss and summarize our results in Sect.~\ref{sec:conclusions}.

\section{The WEAVE quasar mock data}\label{sec:data}

Neural networks fall under the category of supervised learning, requiring labeled datasets of quasar spectra with known absorber information to train and accurately predict new and unseen data. In preparation for analyzing WEAVE data, this work relies on simulated quasar spectra. We anticipate, however, that the model can be deployed with no to minimal modifications on other datasets that share similar characteristics. 
Specifically, for building and profiling the performance of the U-Net, we use $50\,000$ mock WEAVE-like spectra as done in the companion paper by \citet{pistis2025continuum}. The creation of a mock spectrum follows three steps: the generation of the sightline and the absorbers' profiles, the injection of a Ly$\alpha$ forest spectrum, and the addition of noise.

\begin{table}
\centering
\caption{Metal ion names and rest wavelengths for transitions inserted in the mock spectra.}
\label{tab:metal_ions}
\begin{tabular*}{\columnwidth}{l @{\extracolsep{\fill}} l}
\hline\hline
\noalign{\smallskip}
Ion name & Restframe $\lambda_\mathrm{RF}~(\mathrm{\AA}$)\\
\hline
\noalign{\smallskip}
$\ion{O}{i}^{@\dagger}$  & $1302$\\
$\ion{C}{ii}^{@\dagger}$ & $1334$ \\
$\ion{Al}{ii}^{@\dagger}$& $1670$ \\
$\ion{Zn}{ii}^@$         & $2026$ \\
$\ion{Si}{iv}^{@\dagger\circ}$ & $1393, 1402$ \\
$\ion{C}{iv}^{@\dagger\circ}$  & $1548, 1550$ \\
$\ion{Ni}{ii}^@$ & $1709, 1741$ \\
$\ion{Mn}{ii}^@$               & $2576, 2594$ \\
$\ion{Mg}{ii}^{@\dagger\circ}$ & $2796, 2803$ \\
$\ion{S}{ii}^@$ & $1250, 1253, 1259$ \\
$\ion{Fe}{ii}^{@\dagger\circ}$ & $1608, 2382, 2600$ \\
$\ion{Si}{ii}^{@\dagger}$ & $1260, 1304, 1526, 1808$ \\
\hline
\end{tabular*}
\tablefoot{Neutral hydrogen absorption systems class: $^@$DLA (Damped Ly-$\alpha$ Absorbers $\log (N_\ion{H}{i}/{\mathrm{cm}^{-2}})\geq 20.3$), $^\dagger$sub-DLA ($19\leq \log (N_{\ion{H}{i}}/{\mathrm{cm}^{-2}})<20.3$), $^\circ$LLS (Lyman Limit Systems $17\leq \log (N_\ion{H}{i}/{\mathrm{cm}^{-2}})<19$)}
\end{table}

In the first step, a quasar redshift between $2.1<z_\mathrm{QSO}<4.8$ and an observed $r$-magnitude between $16.3<r<21.8$ is assigned to each spectrum. Then, sightlines are populated with strong neutral hydrogen absorbers ($0.2<z_\mathrm{abs}<4.8$ and $17<\log (N_\ion{H}{i}/{\mathrm{cm}^{-2}})<22.5$) and the associated metal absorption lines, depending on the class of the system, as detailed in Table~\ref{tab:metal_ions}.
Physical parameters, distributions, and frequencies of the absorption systems are retrieved from observed data \citep{paris2018dr14, shu2019agngaia, lyke2020dr16sdss}.

Metal absorption line profiles are created using a profile maker based on non-negative matrix factorization \citep[NMF,][]{lee2000nmf}. This NMF profile maker \citep[NMFPM,][]{nmfpm} successfully reproduces realistic line profiles based on physical information, such as the ion producing the line, its rest-frame center, and the value of the column density.
NMFPM can generate a set of noise-free metal profiles at the desired resolution. To avoid sampling effects, the velocity resolution in the NMFPM profiles ($1~\mathrm{km}~\mathrm{s}^{-1}$) is significantly finer than the effective WEAVE velocity sampling at the redshift of the targets, which corresponds to an observed-frame pixel scale value of $18~\mathrm{km}~\mathrm{s}^{-1}$.
For each NMFPM set of lines (i.e., a multiplet), equivalent widths (EWs) and respective errors are determined, and the cumulative distribution functions of the optical depth are computed.
Each multiplet is then inserted into the respective spectrum.
The position of each ion is encoded in a binary label array, with positive values assigned to pixels that encompass the interval $5\%-95\%$ of the optical depth profile of each absorption line.

In the second step, we add low column-density hydrogen lines.  Low column density systems ($\log (N_\ion{H}{i}/{\mathrm{cm}^{-2}})<17$) are statistically represented by the Ly$\alpha$ forest. Ly$\alpha$ forest mocks are created using Ly$\alpha$ Cosmological Lofty Realization \citep[LyaCoLoRe,][]{2023ascl.soft12005F}, a software that uses CoLoRe simulations \citep{Ram_rez_P_rez_2022} to create realistic skewers, $1D$ line-of-sight samples through a $3D$ cosmological volume which reproduce the physics of transmitted flux caused by the forest that shapes quasar absorption spectra.
The forest is added ensuring independence of the absorbers along different sightlines.
As most of these absorbers have associated absorption lines that are weak and buried in the noise of WEAVE-like spectra, for simplicity, we do not track their associated metal content.

In the third step, flux-dependent noise is included to simulate spectra similar to those obtained in WEAVE observations at the observational condition with an air mass $a = 1.107$ and an apparent magnitude of the sky $b_{sky} = 20.92 \,\rm{mag/arcsec^2}$. The simulated spectra have been updated to include atmospheric emission skylines, in contrast to their previous use in \citet{pistis2025continuum}.
The spectra are resampled and convolved with an appropriate line-spread function model to match the low-resolution mode ($R\sim5000$) of the WEAVE survey.
Finally, a continuum shape for the quasar is added. The continuum profiles are reconstructed using principal component analysis from \citet{2011A&A...530A..50P}.
The noise is computed by predicting photon counts from the target and sky, following standard $\mathrm{S}/\mathrm{N}$ calculations based on the instrument specifications.
Using the noise-free model continuum, each spectrum is then normalized to unity over the whole wavelength range.

\section{Absorption line identification with the U-Net architecture}\label{sec:method}

In this work, we aim to develop a new U-Net architecture for detecting metal absorbers in quasar spectra.
In this section, we describe the data pre-processing (Sect.~\ref{subsec:preprocessing}) and the development of the architecture used for the detection (see Sect.~\ref{subsec:abs_det}).
We also briefly describe the autoencoder developed for the continuum fitting by \citet{pistis2025continuum}. This code will be used to assess the stability of the U-Net classifier's performance once uncertainties stemming from the continuum normalization step of quasar spectra are taken into account (see Sect.~\ref{subsec:cont_fit}).

\subsection{Data pre-processing}\label{subsec:preprocessing}

The pre-processing of the data is an important aspect when dealing with NNs \citep[Section 12.2.1]{Goodfellow-et-al-2016}. This step can have a significant impact on the performance of the algorithms used.
For the detection of absorbers, we have considered all generated spectra, regardless of their signal-to-noise ratio $\mathrm{S}/ \mathrm{N}_\mathrm{spectrum}$, computed as the median of the per-pixel signal-to-noise ratios over all the original wavelength ranges. The decision to train on all spectra, as opposed to focusing only on those with higher $\mathrm{S}/ \mathrm{N}_\mathrm{spectrum}$, is discussed in Appendix~\ref{app:selection}.
 
The spectra and flux uncertainties are shifted and linearly interpolated onto a common rest frame wavelength array encompassing the wavelength range $\left[ 1\,230, 3\,095 \right]\mathrm{\AA}$, with spacing of $\sim 0.114\ \mathrm{\AA}$ to maintain the smallest resolution element out of all the observed frames. The upper and lower limits are dictated by the need to include the lower redshift spectra flux and to exclude the Ly$\alpha$ forest in addition to continuum-modeling effects, respectively. We exclude the Ly$\alpha$ forest region due to the blending with hydrogen lines that hinder the detection of metal lines and also due to the presence of a current limitation in the mocks concerning the modeling of the Ly$\alpha$ emission line. In particular, we want to avoid the Ly$\alpha$ broad line emission because we observed that the normalized continuum does not remain at the expected level of unity in this region.
Using a common quasar rest frame ensures fixed wavelength ranges where metal absorption lines appear.

The binary labels are also interpolated over the common wavelength array to align with the positions of the absorption lines. 
During pre-processing of data, wavelengths outside the spectral range (NaN values) are replaced by a flux value of 1 and their corresponding labels are set to 0. 
These regions are not explicitly masked in the loss function; so they are treated as continuum during training.
For a given line, the $\mathrm{S}/ \mathrm{N}_\mathrm{line}$ is defined as its EW over the error on the EW itself: 
\begin{equation}
    \mathrm{S}/ \mathrm{N}_\mathrm{line} \equiv EW / \Delta EW
\label{eq:snr_line}
\end{equation}
EW and its associated error are computed following the derivation in \cite{Vollmann_ews}, assuming that the continuum does not have an error and discretizing over the pixel values to obtain 
\begin{equation}
    EW = \Delta\lambda\displaystyle\sum_{i = \mathrm{px}_i}^{\mathrm{px}_f} \left( 1-f_i \right) \quad \Delta EW = \Delta\lambda\sqrt{\displaystyle\sum_{i = \mathrm{px}_i}^{\mathrm{px}_f}\sigma_{f,i}^2},
\label{eq:snr}
\end{equation}
where $f$ is the normalized flux, $\sigma_f$ the error associated to the flux, $\Delta\lambda$ is the binning spacing of the wavelength array, and $\mathrm{px}_{i,f}$ are the initial and final spectral bins of the lines.
Out of all the injected absorption lines in the spectra, we consider relevant only those with $\mathrm{S}/ \mathrm{N}_\mathrm{line} \geq 1.5$; this threshold is to ensure to take into consideration only potentially detectable absorption lines given the typical quality of a mock spectrum.
The absorption lines with $\mathrm{S}/ \mathrm{N}_\mathrm{line} < 1.5$ are considered undetected from the spectral noise and, as such, their corresponding classification label is set at $0$.
Here and throughout, we adopt the signal-to-noise integrated over the line ($\mathrm{S}/ \mathrm{N}_\mathrm{line}$) when considering detections, as it provides a better metric than the spectrum signal-to-noise ($\mathrm{S}/ \mathrm{N}_\mathrm{spectrum}$)
to determine which absorption lines are detectable because it encompasses both the notion of the error of the flux and the strength of the line itself.
An additional step is required to handle the continuum when using spectra that are not already normalized to unity. To test the effect of residuals relating from imperfect continuum subtraction, we perform the steps of continuum fitting and then the step of line detection.
An additional step before the continuum fitting is to scale all the spectra to the same range to avoid the specific features of some spectra dominating over others.
Each quasar spectrum is normalized by the median flux at ${\lambda = 1\,450\,\mathrm{\AA}}$ within a window of $\Delta \lambda = 50\,\mathrm{\AA}$.

\subsection{The U-Net architecture: a model for absorber detection}\label{subsec:abs_det}

The U-Net model, first described by \citet{ronneberger2015unet}, has a convolutional NN architecture characterized by an encoder, a symmetrical decoder path, and a bottleneck linking the two.
The encoder progressively decreases spatial resolution, forcing the U-Net to learn a compressed representation of the image. In contrast, the decoder path increases the spatial resolution, reconstructing the image from its representation.
The bottleneck, as the name suggests, is the bottom narrowest part of the network, connecting the encoder and the decoder.
An architecture that consists of an encoder and a decoder are known as autoencoders \citep{2006Sci...313..504H}; however, they often suffer from poor location accuracy.
The U-Net, instead, employs skip connections, concatenating the higher resolution features from the encoder to the upsampled features of the decoder, propagating information to higher resolution layers. Skip connections help to preserve fine details of the spectrum while incorporating high-level features captured in deeper layers, preserving the spatial localization of features of the image.
In a classification task, such as that of detecting metal absorption lines \citep{zhao_mgii, xia_caii, Szakacs_mgii}, the choice of using a U-Net architecture is supported by the encoder-decoder structure capable of high location accuracy without abandoning the use of the surrounding context.

Despite the rapid emergence of transformer and attention-based architectures capable of spectral feature extraction \citep[e.g.,][]{Koblischke, Fortino2025ABCSNAC}, the use of a U-Net model is preferred for precise segmentation.
The U-Net architecture provides a strong spatially-located inductive bias that enables the precise localization of narrow absorption features such as those imprinted by metal ions.

As the model handles spectra, the input is a 1D image with a length equal to the number of pixels in the common rest-frame defined above. For each pixel, the model predicts whether it belongs to a metal absorption line, operating in a binary classification.
The input spectrum is processed in the first block of the encoder path, which consists of a double convolution immediately followed by a max pooling operation with a fixed pool size and stride of 2. In the convolutions, the feature maps are padded to maintain their spatial dimensions across layers. The operations of the first block are repeated for three layers. Between each block, a dropout layer is inserted to prevent overfitting.
At the bottom of the model, there is the bottleneck path that links the encoder and decoder. The bottleneck layer consists of a double convolution on the most condensed (low-resolution) spatial representation of the feature maps.

The decoder is symmetrical with respect to the encoder, with transposed convolution layers in place of max pooling layers. At each block, it doubles the spatial size and maintains the same number of filters as the corresponding encoder level, mirroring the trend of the encoder path.
After every step of the encoder path, skip connections concatenate feature maps at the same resolution level at corresponding layers between the encoder and decoder.
To restore the original input dimension, the final operation is a convolution with a linear activation function, where the number of filters is reduced to match the number of classes the model is trying to predict. In the case of a binary classification, this means reducing the output to $2$ filters, each with a kernel size of $1$.

\begin{table}
\centering
\caption{Hyperparameters of the optimized U-Net for line identification.}
\label{tab:hyper_unet}
\begin{tabular*}{\columnwidth}{l @{\extracolsep{\fill}} l}
\hline\hline
\noalign{\smallskip}
Hyperparameter & Value\\
\hline
\noalign{\smallskip}
Base size & $4$\\
Increment $1$ & $1$\\
Increment $2$ & $0$\\
Increment $3$ & $1$\\
Dropout rate & $0$\\
Weight & $2$\\
Kernel size & $5$\\
\hline
\end{tabular*}
\end{table}

\begin{figure*}
    \resizebox{\hsize}{!}{\includegraphics[width=1.0\textwidth]{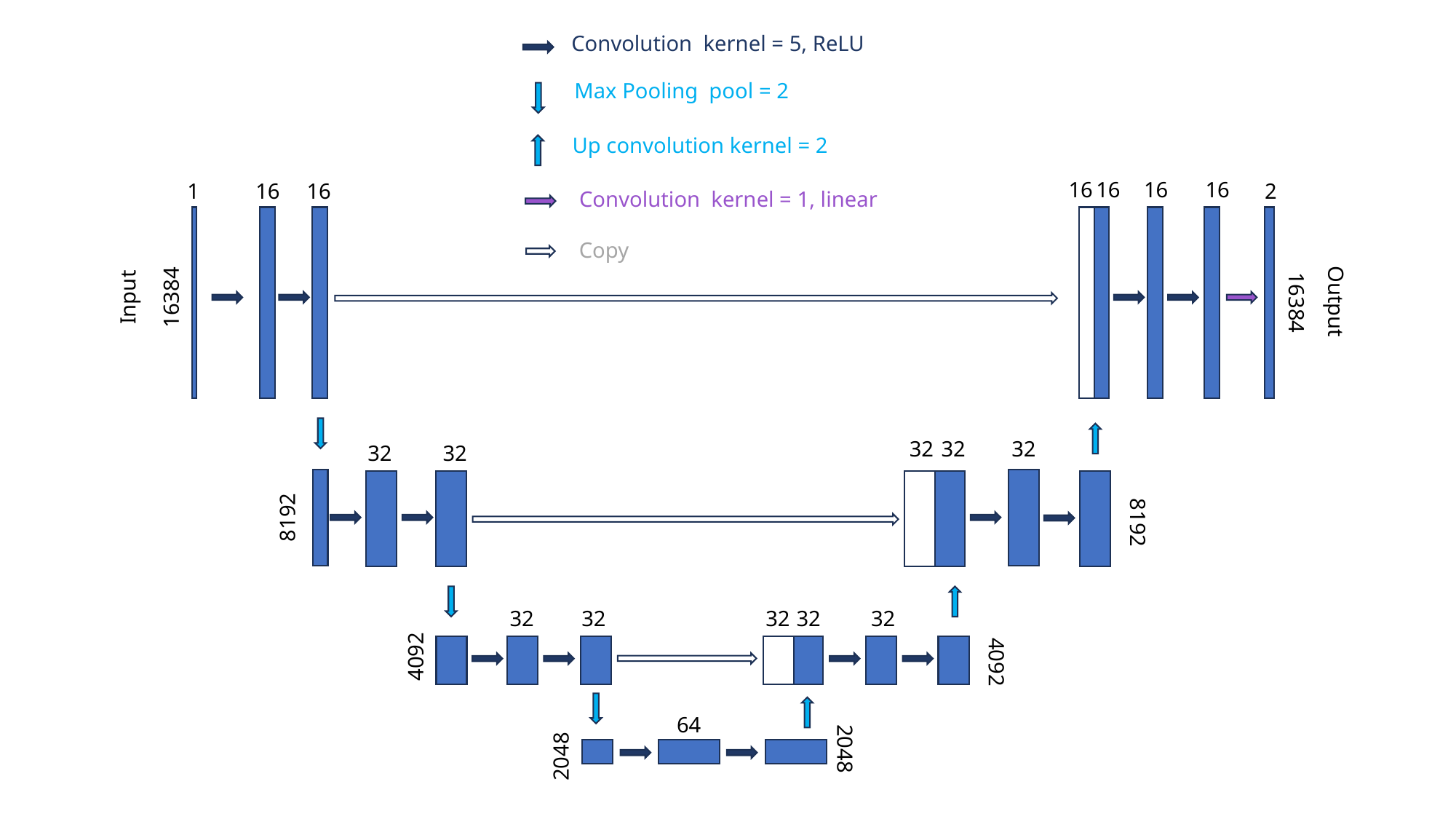}}
    \caption{Scheme of the architecture of the optimized U-Net model applied to WEAVE mock data for metal lines detection. The blocks are the different layers in the U-Net, with the white blocks being the ones copied by the skip connections. The different colored arrows represent operations between layers. The activation function used is, in all cases, the ReLU function. The number above the blocks denotes the number of feature maps in the layer, while the number on the side is the size of each layer.}
    \label{fig:unet}
\end{figure*}
\begin{figure}
    \resizebox{\hsize}{!}{\includegraphics[width=1.0\columnwidth]{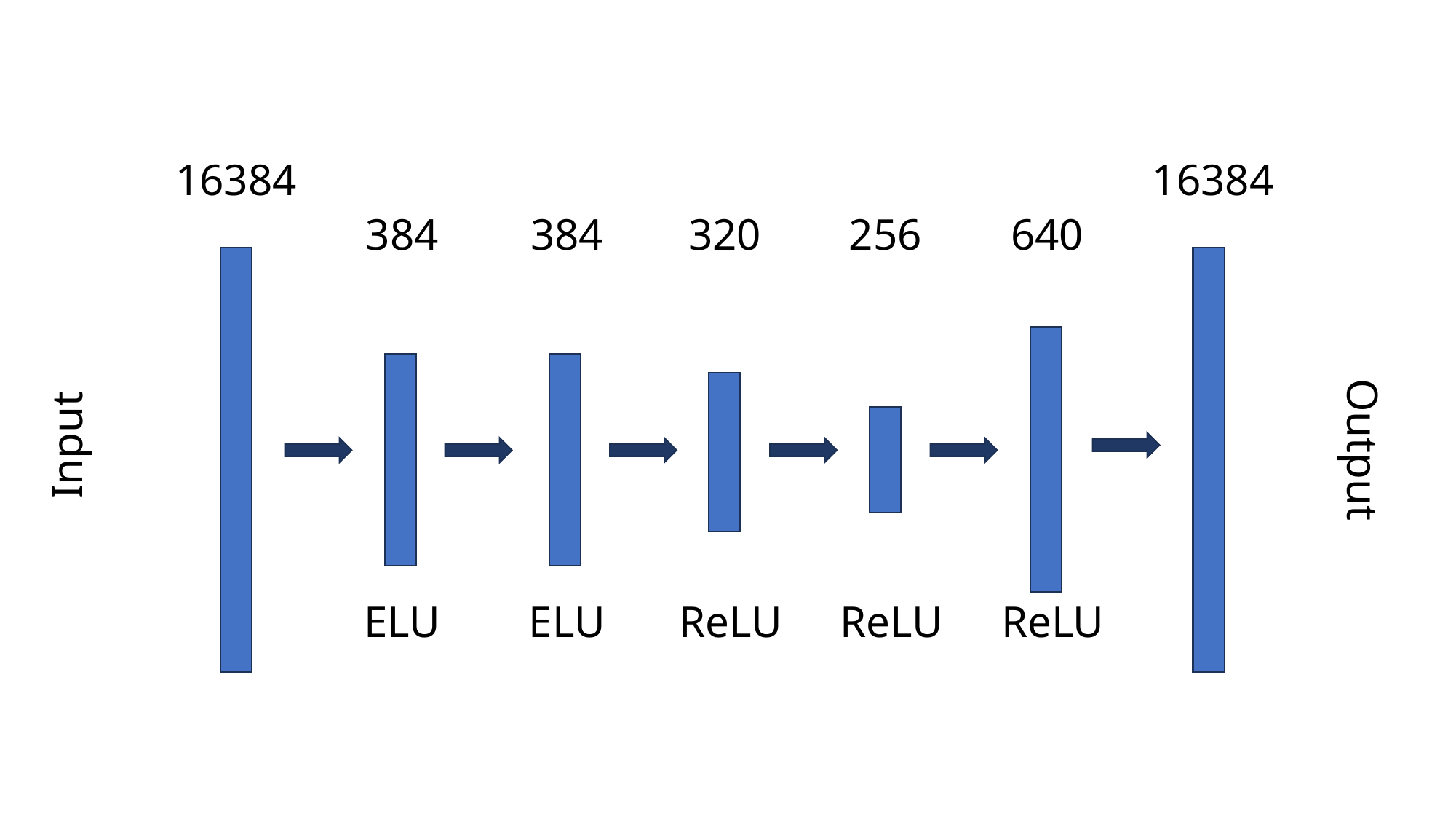}}
    \caption{Scheme of the architecture of the optimized autoencoder model applied to WEAVE mock data for the continuum estimation. The blocks are the different layers in the autoencoder. The number above the blocks denotes the size of the layer. The activation function used between each layer is written below the blocks.} All the layers represented are dense layers.
    \label{fig:autoencoder}
\end{figure}

The specific choices in building the architecture come from a hyperparameter optimization done using the Bayesian optimization implemented within Keras Tuner 1.4.7 \citep[][]{omalley2019kerastuner} on $50$ trials, aiming for the lowest validation loss value.
The search space encompasses the number of filters present in each layer, expressed as a power of 2, starting from the power of the base size and incrementing cumulatively at each encoder level.
Since the decoder path is symmetrical with the encoder path, it has a mirrored number of filters.
Additional hyperparameters include the kernel size and dropout layer probability, which are fixed for all blocks.
A final hyperparameter is the class weight assigned to positive labels in the loss function. Since the metal line features are underrepresented compared to continuum values, we use weights to penalize the loss more in case of a misdetection of a metal line.
The loss function used is a weighted sparse categorical cross-entropy function:
\begin{equation}
    L = -w_c\, log(\hat{y_c}),
\end{equation}
where $\hat{y_c}$ is the probability of the predicted label and $w_c$ the class weight for the true class.
The values of the hyperparameters of the best model are summarized in Table~\ref{tab:hyper_unet}.
A few interesting observations can be made on the results of the hyperparameter optimization.
The dropout rate probability of $0$ indicates that all nodes are considered essential, and therefore, regularization via node dropout does not play a significant role. The second increment is $0$, suggesting that it is not necessary to increase the number of filters to better learn representation at a higher level of abstraction in the deepest layers. Ultimately, the weight assigned in the loss function is $2$, which is the lowest admissible value, since a weight of $1$ would correspond to equal importance being assigned to both classes. This means that the model can learn the absorption features without requiring heavy penalization in the loss function.
The final U-Net architecture constructed with these hyperparameters is described in Fig.~\ref{fig:unet}.

The $50\,000$ mock spectra, created as explained in Sect.~\ref{sec:data}, are divided in a $70\%$--$15\%$--$15\%$ proportion for train, validation, and test sets.
To improve training efficiency the $\sim35\,000$ spectra of the train data set were processed with a batch size of $32$ and pre-fetched, meaning that the model processed the data taking $32$ spectra at a time and prepared next batches of data in advance. For each set of hyperparameters, the model was trained for $1000$ epochs. This large number allowed the model to train indefinitely until an early stopping based on the validation loss, using a patience of $5$ epochs and automatic restoration of the best weights. The $\sim7\,500$ spectra of the validation data set are used to evaluate the performance during training. Lastly, the $\sim7\,500$ test spectra are used to assess the model performance based on the metrics described in Sect.~\ref{sec:metrics}.

\subsection{Autoencoder for continuum fitting}\label{subsec:auto}

Mock data have the artificial, yet desirable, property that the continuum level of the quasars is known with no uncertainty. This valuable feature of the mock library enables us to develop the U-Net and assess its intrinsic performance without confusion arising from artifacts of inexact continuum estimation. However, we are also interested in assessing the purity and completeness of absorption line recovery in real-life applications, where the true continuum is unknown, and we thus repeat the study of the performance of the network starting on mocks with a realistic quasar continuum, to which we apply a normalization step. 

\begin{table}
\centering
\caption{Hyperparameters of the optimized autoencoder for continuum normalization.}
\label{tab:hyper_auto}
\begin{tabular*}{\columnwidth}{l @{\extracolsep{\fill}} l}
\hline\hline
\noalign{\smallskip}
Hyperparameter & Value\\
\hline
\noalign{\smallskip}
Number of layers & $2$\\
Size encoder $1$  & $384$\\
Activation encoder $1$ & ELU\\
Size encoder $2$  & $384$\\
Activation encoder $1$ & ELU\\
Size bottleneck & $320$\\
Activation bottleneck & ReLU\\
Size decoder $1$  & $256$\\
Activation decoder $1$ & ReLU\\
Size decoder $2$  & $640$\\
Activation decoder $2$ & ReLU\\
Learning rate & $10^{-4}$\\
\hline
\end{tabular*}
\end{table}

The continuum normalization step was performed on the WEAVE mocks using an autoencoder. The choice of using an autoencoder is supported by the results of \citet{pistis2025continuum}, in which an autoencoder outperformed a CNN and a U-Net in predicting quasar continua.
We optimized that autoencoder architecture for our problem. In fact, unlike previous works \citep{liu2021quasar, turner2024lycan, pistis2025continuum}, we do not limit the spectra in the restframe wavelength range ${1\,020 < \lambda\ (\mathrm{\AA}) < 2\,000}$, but we modify the wavelength range considered to that of the mock spectra ($\left[ 1\,230, 3\,095 \right]\mathrm{\AA}$). This, combined with the chosen smaller spectral resolution with respect to the previous work \citep{pistis2025continuum}, meant we needed to train a new autoencoder.

The autoencoder has a classical architecture, which includes a part that compresses the input spectra into a smaller representation, the encoder, and a part that reconstructs the original input from that, the decoder. A masking layer allows the autoencoder to ignore NaN values of the spectra.
During the optimization step, the size of each layer was not constrained to the size of the previous layer.
This choice led to an unconventional autoencoder where the encoder and the decoder have asymmetrical sizes.

Each layer could vary its dimension between $128$ and $1024$, with a step of $128$.
Additionally, the activation function is left free to change between layers, allowing it to be either a rectified linear unit function \citep[ReLU,][]{ReLU} or an exponential linear unit function \citep[ELU,][]{ELU}.
Between the encoder and the decoder, there is a bottleneck layer.
A last hyperparameter is the learning rate of the model, chosen between $10^{-2},\,10^{-3},\,10^{-4}$.
We perform a hyperparameter optimization using a random search tuner on $50$ trials, aiming for the lowest validation loss value, with a loss function based on the mean squared error. In Table~\ref{tab:hyper_auto}, the optimized values obtained from the hyperparameter search are summarized.
The specifics of the model's architecture are illustrated in Fig.~\ref{fig:autoencoder}.

\subsection{Metrics for assessing the network performance}\label{sec:metrics}

We used different metrics to quantify the performance of our NNs according to their respective goals.
For the detection of absorbers, we used the completeness,
defined as the fraction of absorption lines retrieved among all the ones with $\mathrm{S}/ \mathrm{N}_\mathrm{line} \geq 1.5$, and purity, defined as the fraction of the real lines among all the retrieved ones.
Completeness (C) and purity (P) can be expressed as
\begin{equation}
    \text{Completeness} =\frac{\mathrm{TP}}{\mathrm{TP} + \mathrm{FN}},
\end{equation}
and
\begin{equation}
    \text{Purity} =\frac{\mathrm{TP}}{\mathrm{TP} + \mathrm{FP}}
\end{equation}
where $\mathrm{TP}$ are the true positive detections, $\mathrm{FN}$ the false negative ones, and $\mathrm{FP}$ the false positive ones.

The U-Net model described in Sect.~\ref{subsec:abs_det} outputs an array of labels with the exact spatial dimensions as the input spectrum, meaning that the predictions are made pixel by pixel.
Given that the ultimate goal is to detect absorption lines, the evaluation of the model through metrics computation should focus on the detection of entire lines, rather than individual pixels.
For this reason, a true positive detection is recorded when at least one predicted positive pixel coincides with the corresponding positions of the ground truth labels of an absorption line.
Otherwise, the detection is classified as a false negative.
The U-Net model might predict a common label for close absorptions or blended lines. Since the metric calculations are based on the true and predicted labels and we calculate a single centroid (see Sect.~\ref{subsect:position} for the error on the position of the line centroids) for each predicted absorption, in both cases we count TPs for all the multiple lines and blended lines.
Similarly, a false positive is defined as a predicted detection that does not correspond to a system in the ground truth labels.
Additionally, to provide a more balanced evaluation of the U-Net's performance, accounting for both completeness and purity, we used the F1 score metric:
\begin{equation}
    \text{F1 score}= 2\;\cdot\;\frac{\mathrm{Completeness}\;\cdot\;\mathrm{Purity}}{\mathrm{Completeness}\;+\;\mathrm{Purity}}\:.
\end{equation}
True metal absorption lines not completely within the boundaries of the spectra are not considered as lines in the metric computation.

To assess the quality of the continuum estimation, we used the absolute fractional flux error \citep[AFFE; e.g.,][]{liu2021quasar, turner2024lycan, pistis2025continuum} defined as:
\begin{equation}
    \left| \delta F \right| = \left. \int_{\lambda_1}^{\lambda_2} \left| \frac{F_\mathrm{pred}\left( \lambda \right) - F_\mathrm{true}\left( \lambda \right)}{F_\mathrm{true}\left( \lambda \right)} \right|\, d\lambda \middle/ \int_{\lambda_1}^{\lambda_2} d\lambda \right.,
\end{equation}
where $F_\mathrm{pred}$ is the predicted output and $F_\mathrm{true}$ is the true continuum of the simulated quasar spectra.

\section{Application to the WEAVE mock data}\label{sec:application}

From the $50\,000$ mock spectra, the train and validation spectra are used to optimize the U-Net model as described in Sect.~\ref{subsec:abs_det} and to optimize the autoencoder for continuum fitting as described in Sect.~\ref{subsec:auto}.

The optimized models are then applied to the test subset to evaluate their performance.
Firstly, we evaluated the U-Net performance on the spectra with the true underlying continuum (see Sect.~\ref{subsec:metal_det} and Sect.~\ref{subsect:position}). Then, to simulate the detection process starting from real observed data, the U-Net model is tested on the same test spectra, but with the continuum produced by the autoencoder (see Sect.~\ref{subsec:cont_fit}).
The training and testing of the U-Net and autoencoder models were performed on three NVIDIA A2 GPUs, each with 15 GB of memory, and were implemented using Python 3.12.0 and the open-source machine learning libraries Keras 3.9.2 \citep{keras} and TensorFlow 2.19.0 \citep{tensorflow}.
Application of the autoencoder and of the U-Net on the $\sim7500$ test spectra took $\sim2.7\rm{sec}$ and $\sim21\rm{sec}$ on $1$ NVIDIA A2 GPU, respectively.

\subsection{Detection of metal lines}\label{subsec:metal_det}

\begin{figure}
    \centering
    \includegraphics[width=\columnwidth]{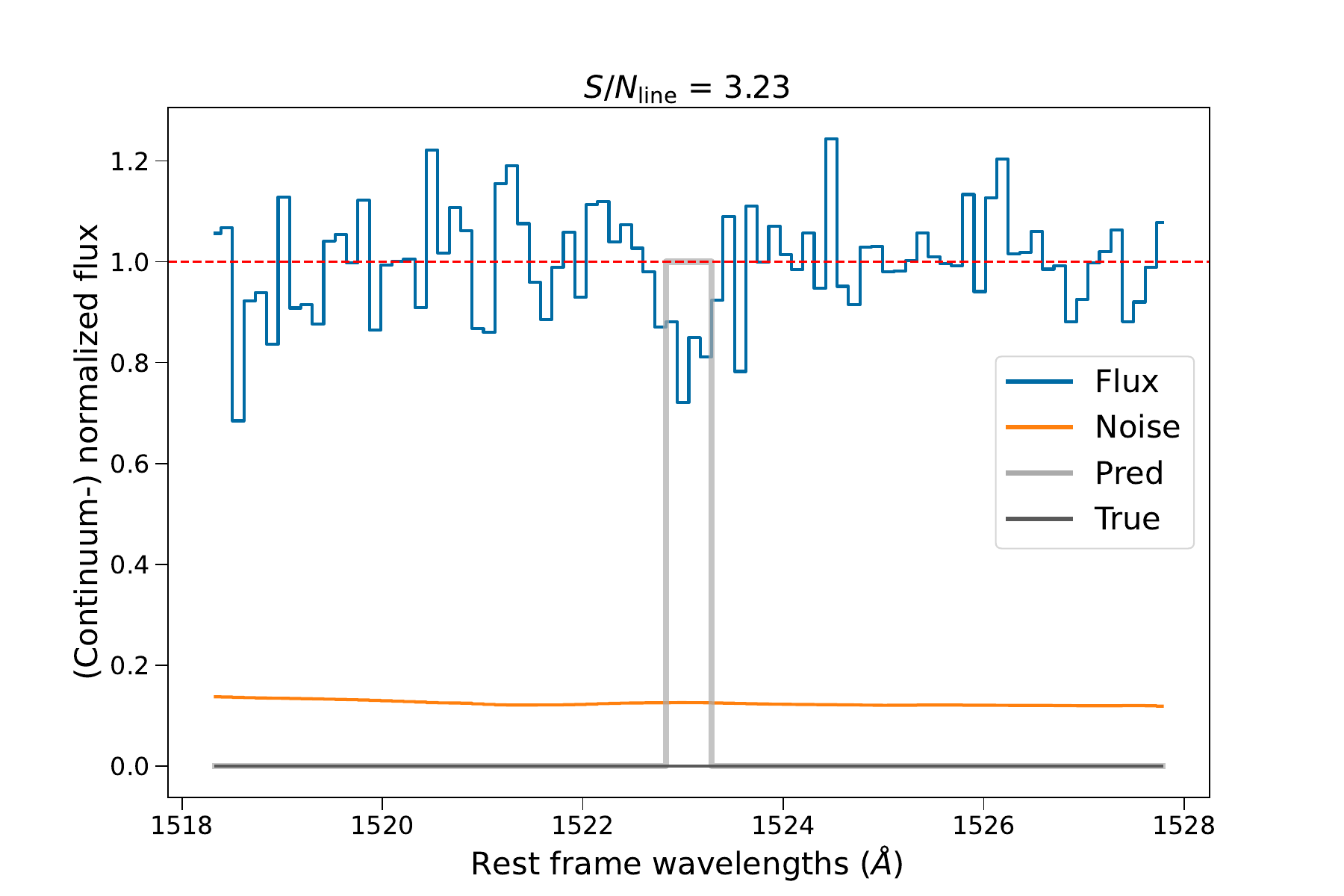}
    \includegraphics[width=\columnwidth]{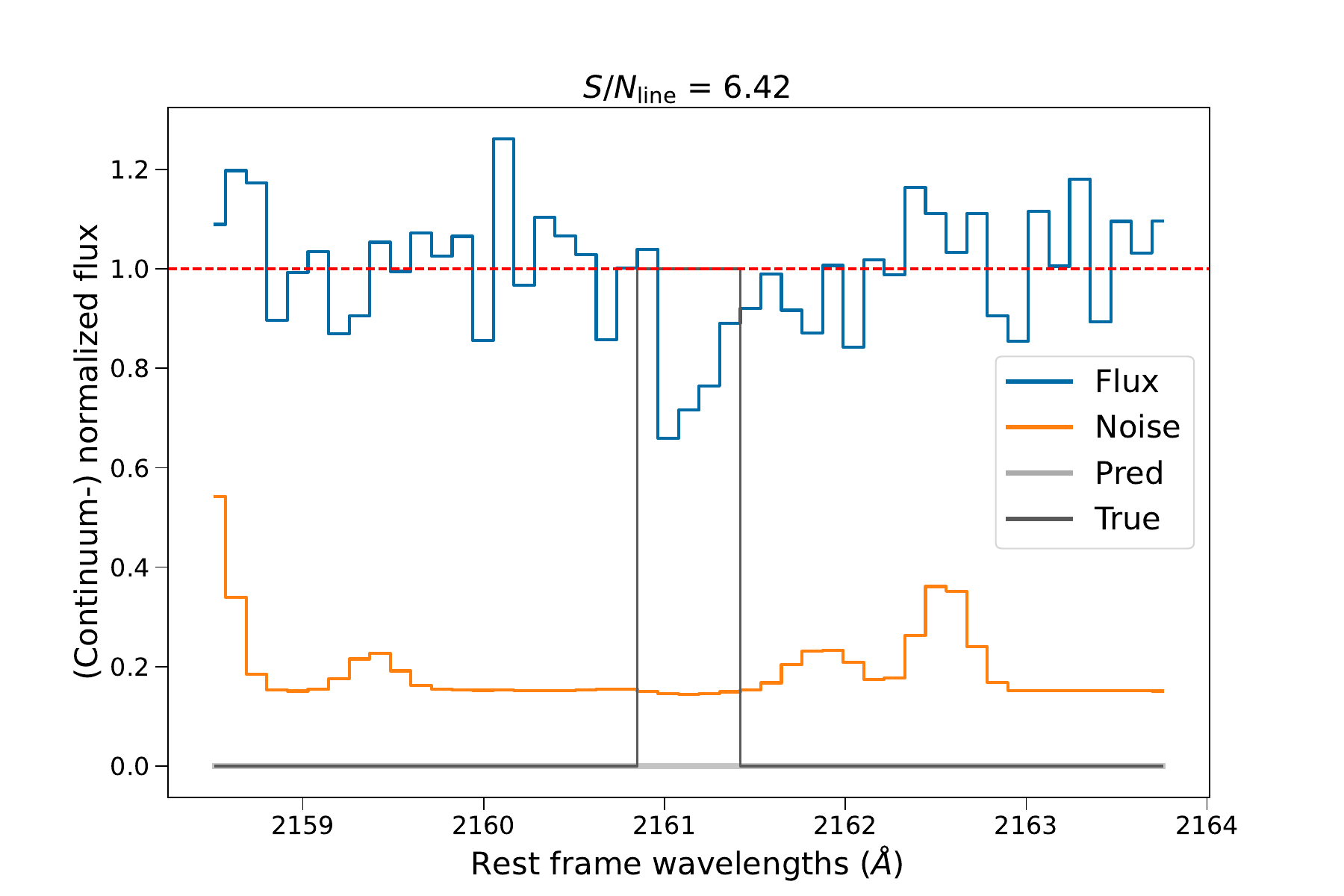}
    \caption{Examples of false positive detection with $\mathrm{S}/ \mathrm{N}_\mathrm{line}=3.23$ (top) and a false negative detection of a line with $\mathrm{S}/ \mathrm{N}_\mathrm{line}=6.42$ (bottom). The spectrum (blue) is normalized to the continuum, with the dashed red line marking unity. Orange line represents noise, light gray the predicted labels, and dark gray the true labels.}
    \label{fig:examples_FN_FP}
\end{figure}

\begin{figure}
    \includegraphics[width=\columnwidth]{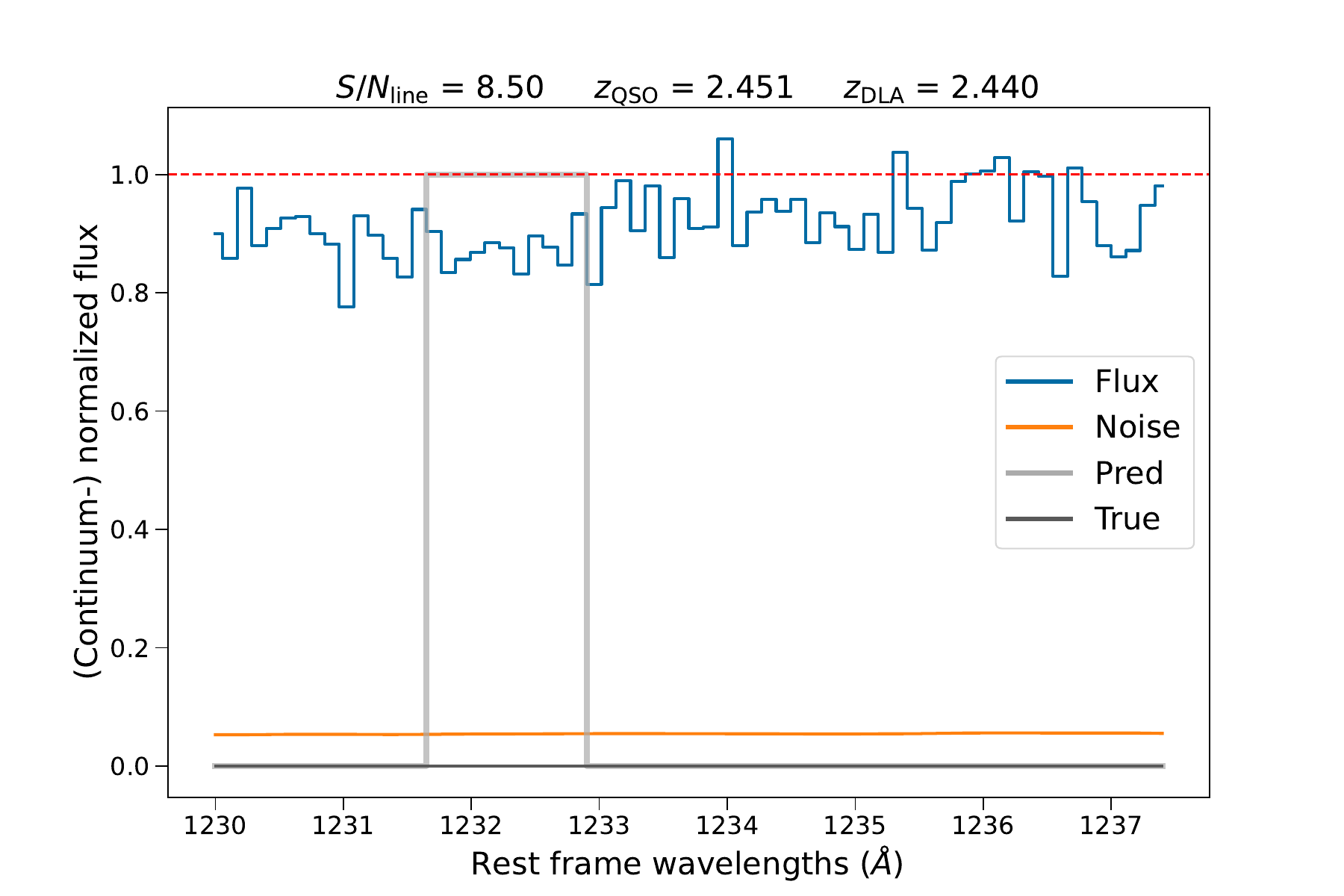}
    \caption{Example of a false positive detection caused by the presence of a DLA absorption feature at $z_{dla}=2.440$ for a quasar at $z_{qso}=2.451$. The presence of the DLA is apparent from the average continuum level is below unity (red dashed line). The conventions of the lines are the same as Fig. \ref{fig:examples_FN_FP}.}
    \label{fig:spectra_examples}
\end{figure}

\begin{figure}
    \centering
    \includegraphics[width=\columnwidth]{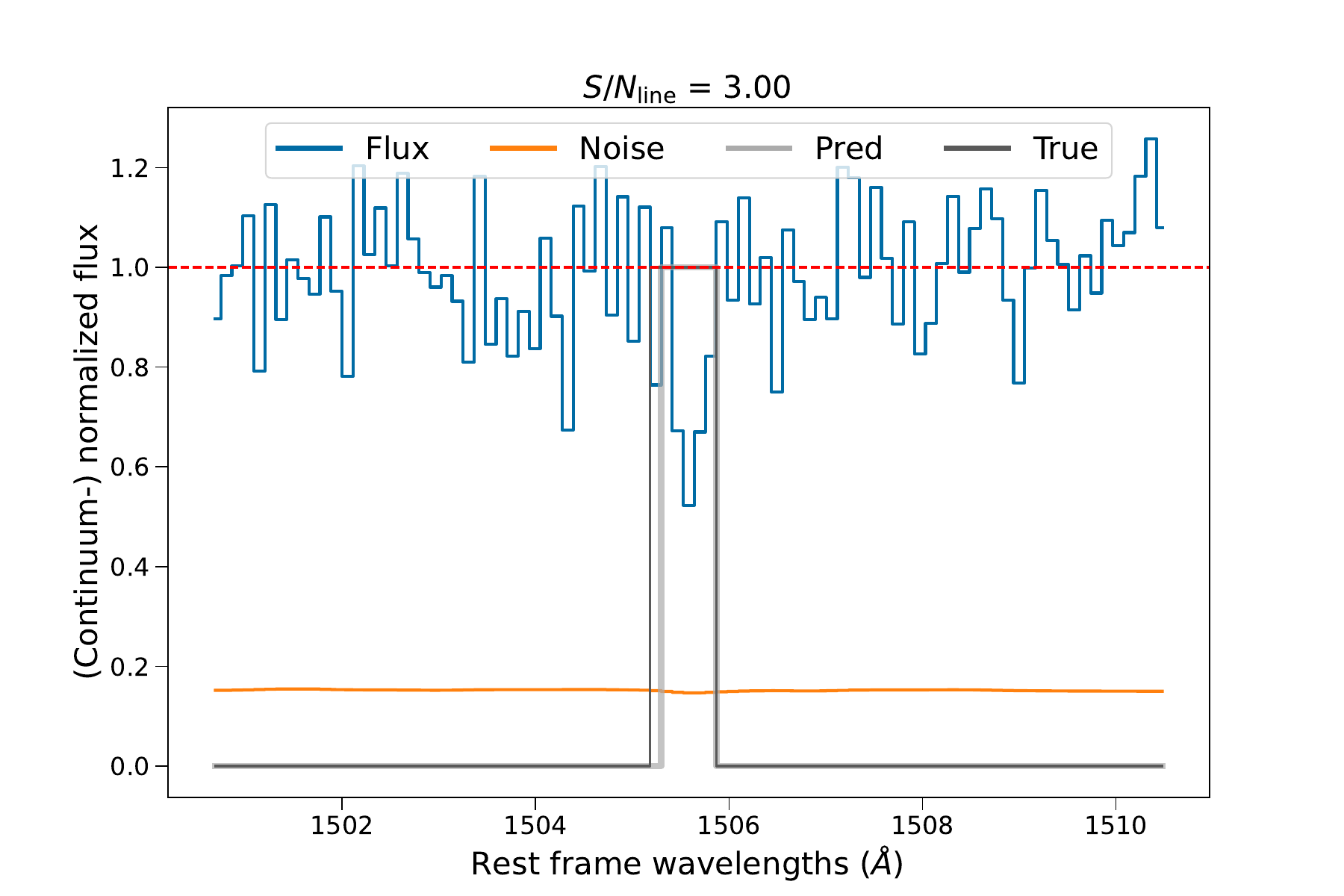}
    \includegraphics[width=\columnwidth]{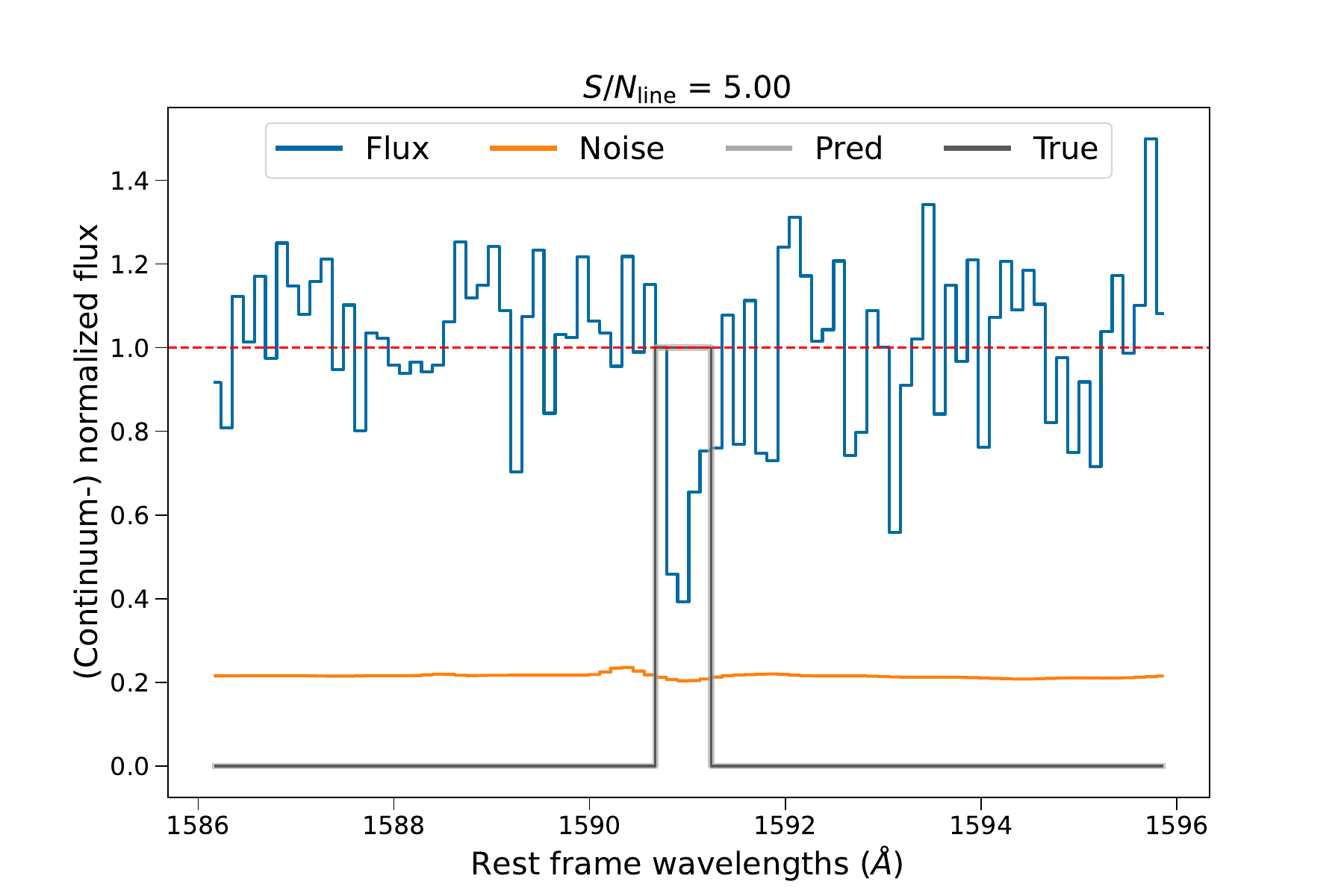}
    \includegraphics[width=\columnwidth]{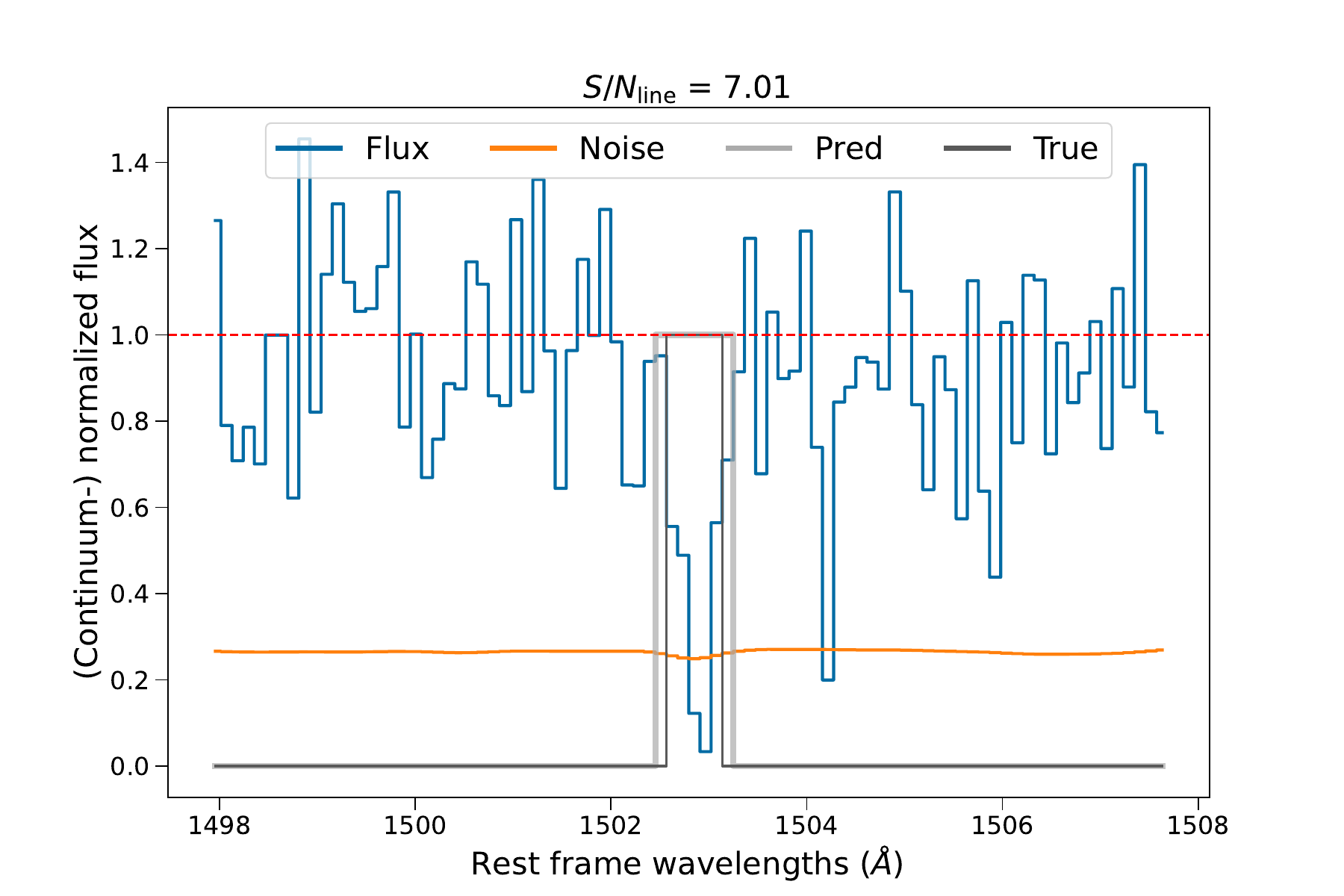}
    \caption{Examples of true positive detections of lines with $\mathrm{S}/ \mathrm{N}_\mathrm{line}=3$ (top), $\mathrm{S}/ \mathrm{N}_\mathrm{line}=5$ (center), $\mathrm{S}/ \mathrm{N}_\mathrm{line}=7.01$ (bottom). The conventions of the lines are the same as Fig. \ref{fig:examples_FN_FP}.}
    \label{fig:line_snr_examples}
\end{figure}

We applied the optimized U-Net model to the test spectra, predicting pixel-wise labels for binary detection of the absorbers.
The metrics described in Sect.~\ref{sec:metrics} are employed to determine the effectiveness of the model.
Focusing on only the predictions within the wavelength range of the spectra, we then computed the number of TP, FP, and FN. 
For each detection, real or false, we study the distribution of the metrics as a function of the signal-to-noise ratio ($\mathrm{S}/ \mathrm{N}_\mathrm{line}$) of the line.
For a TP or FN detection, the $\mathrm{S}/ \mathrm{N}_\mathrm{line}$ is the one given in the spectra creation. 
For a FP detection, instead, the $\mathrm{S}/ \mathrm{N}_\mathrm{line}$ is computed on the predicted positive pixels group, always using Equation~\ref{eq:snr}.
Therefore, the metrics used in this work are simulation-based diagnostics and the actual values for an observational catalog might vary when using a uniform measurement of the$\mathrm{S}/ \mathrm{N}_\mathrm{line}$.
In Fig.~\ref{fig:examples_FN_FP} there are displayed examples of a false positive detection and a false negative missed by the model.
\begin{figure*}
    \resizebox{\hsize}{!}{\includegraphics[width=1.0\textwidth]{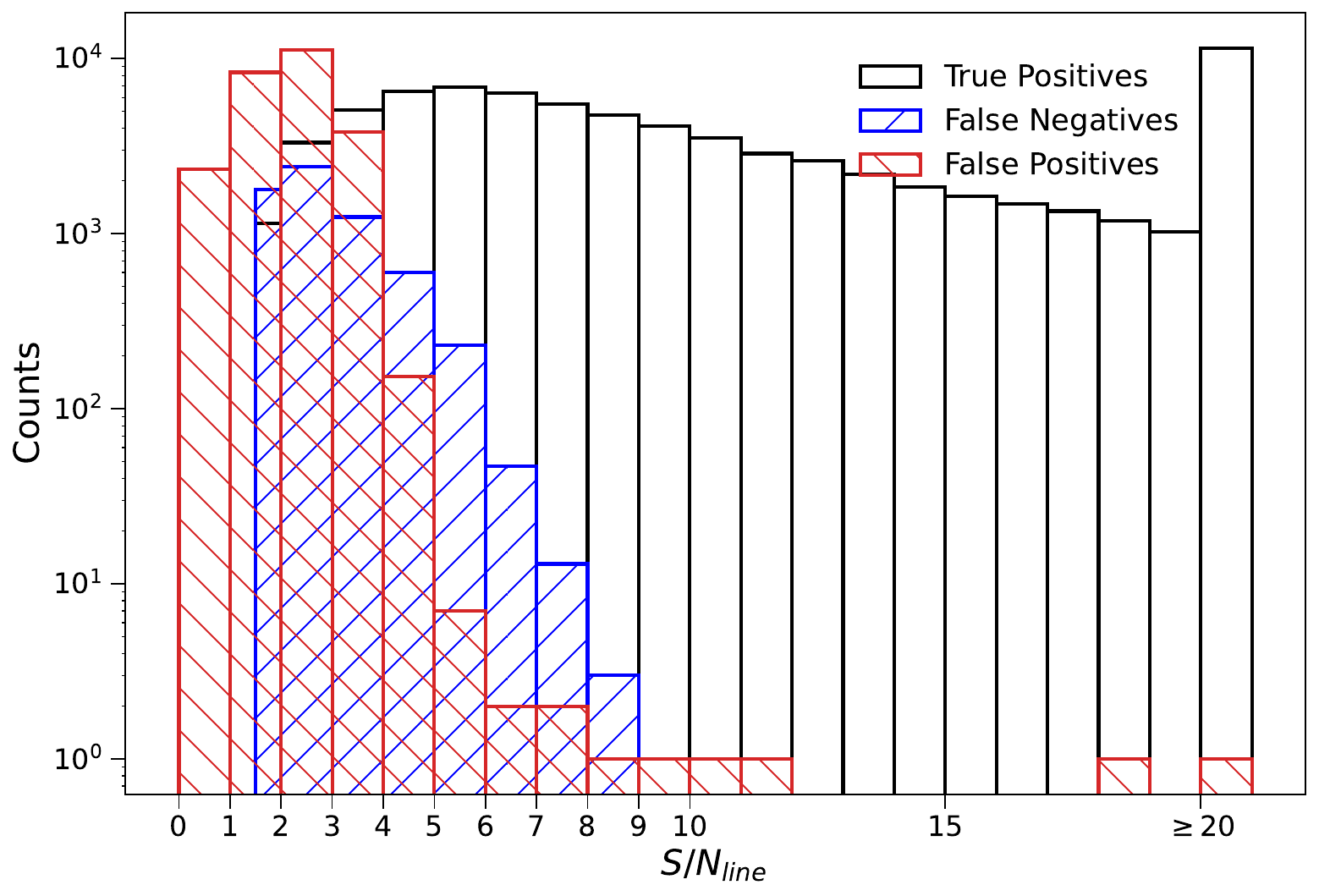}\includegraphics[width=1.0\textwidth]{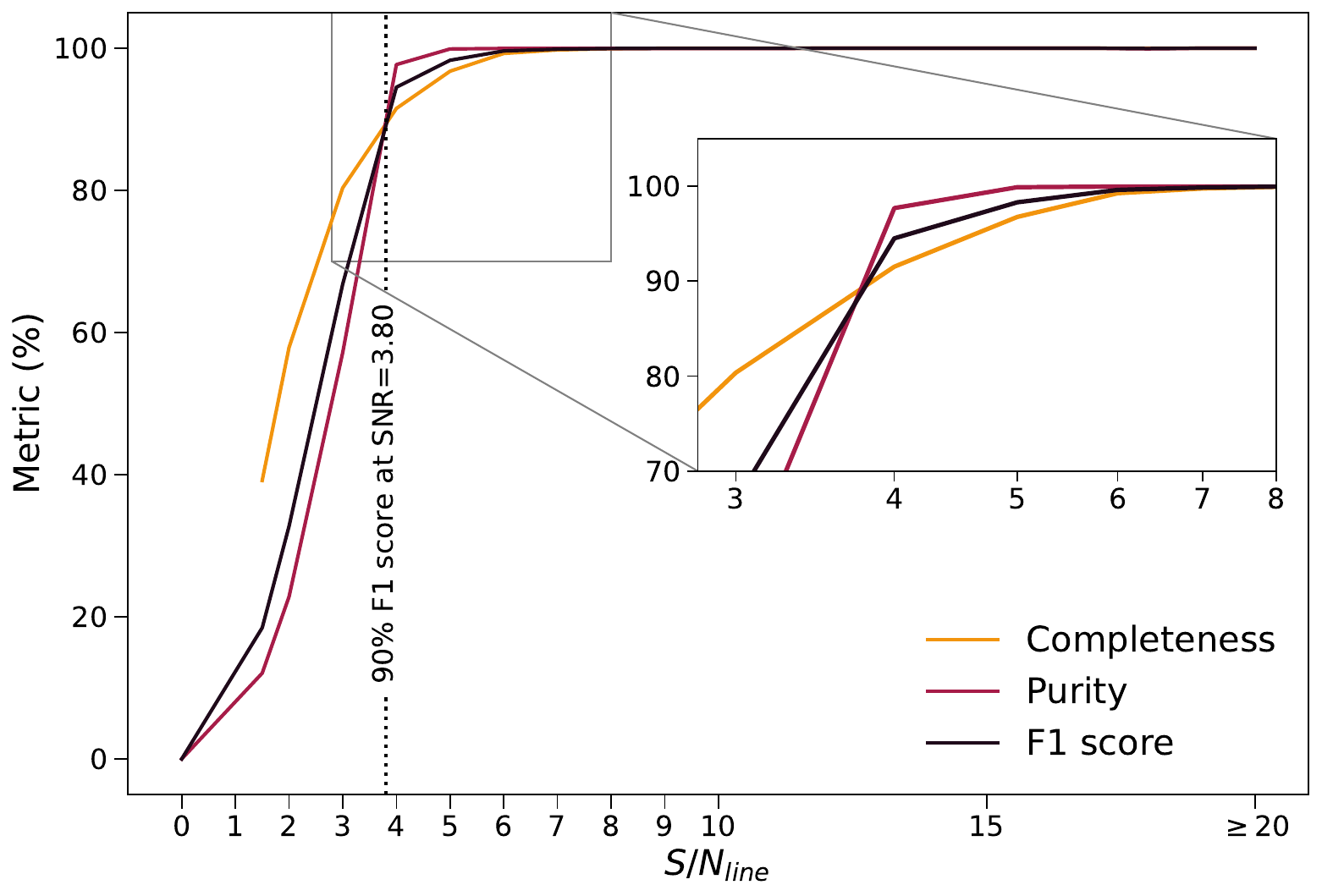}}
    \caption{Performance of the U-Net for the line detection as a function of the $\mathrm{S}/\mathrm{N}_\mathrm{line}$. Left panel: Histograms for true positives (black), false positives (red), and false negatives (blue) in $\mathrm{S}/\mathrm{N}_\mathrm{line}$ bins. The last bin is cumulative for $\mathrm{S}/\mathrm{N}_\mathrm{line} \geq 20$. Right panel: curves of different metrics (completeness in yellow, purity in dark red, and F1 score in black) as a function of the $\mathrm{S}/\mathrm{N}_\mathrm{line}$ . The dotted line highlights the $\mathrm{S}/\mathrm{N}_\mathrm{line}$ level at which the F1 score reaches $90\%$.}
    \label{fig:det_metrics}
\end{figure*}
Figure~\ref{fig:det_metrics} shows the performance of the U-Net. In the left panel, a histogram shows that FNs and FPs are found typically at $\mathrm{S}/ \mathrm{N}_\mathrm{line} \leq 5$, except for a tail that extends at higher signal-to-noise.
Based on a visual inspection of these $17$ more extreme cases of FPs detections with $\mathrm{S}/ \mathrm{N}_\mathrm{line} > 5$, all of these are located near DLA absorption features in the spectrum.
When DLA absorption wings are present, they are often the cause of false positive detections, also for lower $\mathrm{S}/ \mathrm{N}_\mathrm{line}$ lines, due to a drop in the continuum that leads to misdetections.
For all FPs with $\mathrm{S}/ \mathrm{N}_\mathrm{line} > 5$, the redshifts of the DLAs present in the spectra are close to the redshift of the quasar, with an average of $\Delta z=z_{qso}-z_{dla}\simeq0.015$.
An example of this effect is shown in Fig.~\ref{fig:spectra_examples}, where the false positive detection is due to the presence of a DLA wing, recognizable thanks to the continuum level that is consistently below unity.
Appropriate identification and either masking or subtraction of DLAs \citep[using, e.g.,][]{Prochaska_dla, Noterdaeme_dla, Garnett_dla} is thus recommended as a further step.
The observed decrease in TP at higher $\mathrm{S}/ \mathrm{N}_\mathrm{line}$ is, instead, due to the fact that there are fewer high $\mathrm{S}/ \mathrm{N}_\mathrm{line}$  lines available to detect. 
The right panel of Fig.~\ref{fig:det_metrics} shows both completeness and purity reaching a $90\%$ level around $\mathrm{S}/ \mathrm{N}_\mathrm{line} \approx 4$, with the F1 score reaching $90\%$ at $\mathrm{S}/\mathrm{N}_\mathrm{line} = 3.8$.
The model is able to perform with high completeness and purity not only in the high $\mathrm{S}/ \mathrm{N}_\mathrm{line}$ regime ($\mathrm{S}/ \mathrm{N}_\mathrm{line} \geq 8$, where all the metrics reach $100\%$) but also at lower values. At $\mathrm{S}/ \mathrm{N}_\mathrm{line}\sim4$, the model is still performing at satisfactory levels (metrics$\sim90\%$). These high values of completeness, purity, and F1 score indicate that the model successfully identifies the most relevant samples and that most of the identified samples are accurate. For lines with $\mathrm{S}/ \mathrm{N}_\mathrm{line}\lesssim4$ the performance drops. The model is nevertheless able to detect a large number of lines; however, the purity declines and it is hard to make use of this low $\mathrm{S}/ \mathrm{N}_\mathrm{line}$ sample.
Figure~\ref{fig:line_snr_examples} shows examples of true positive detections of lines with different $\mathrm{S}/ \mathrm{N}_\mathrm{line}$.

\begin{figure}
    \resizebox{\hsize}{!}{\includegraphics[width=1.1\textwidth]{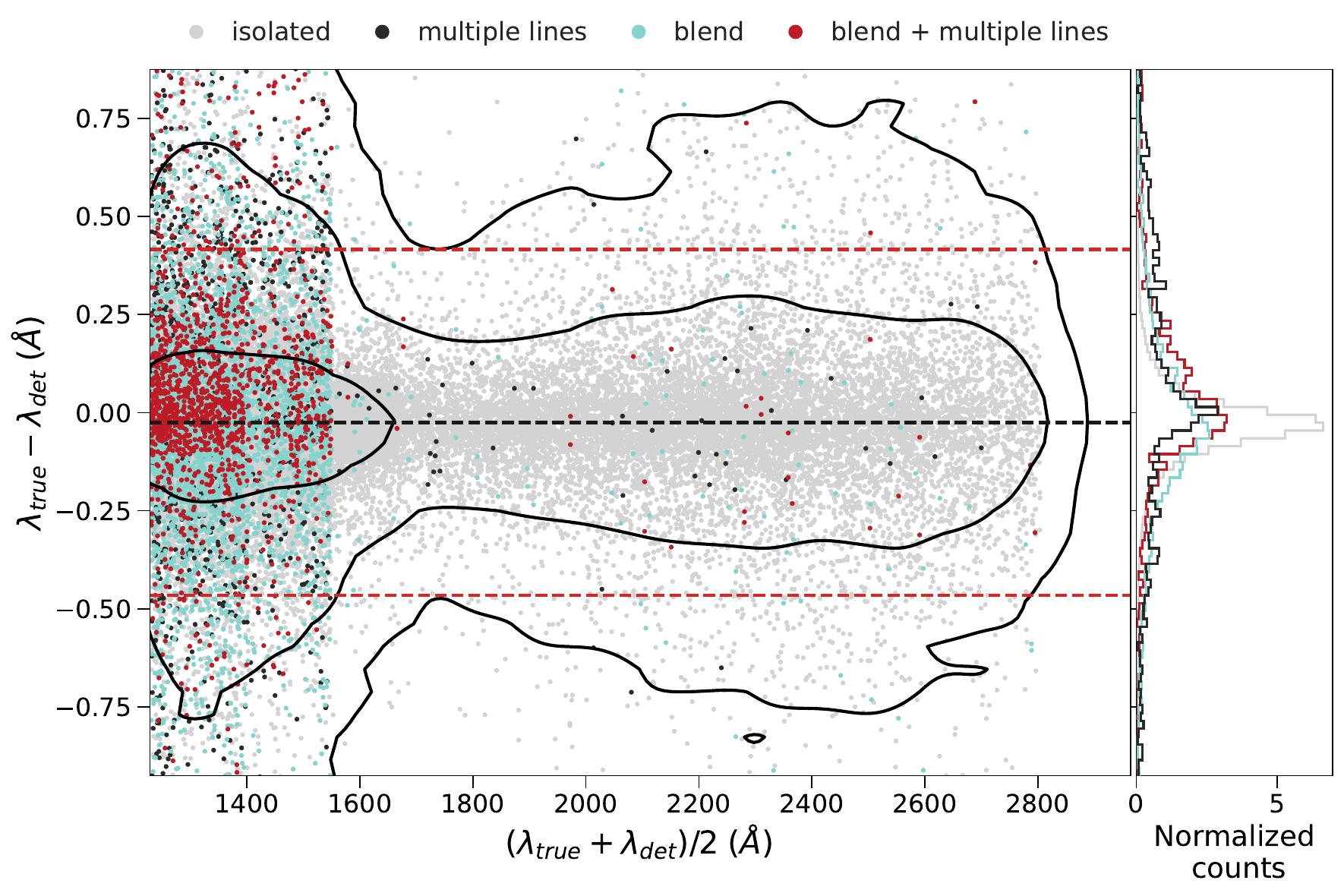}}
    \caption{Bland–Altman plot for the true and predicted centers of the absorption lines. The mean difference is marked by a dashed black line, and the $95\%$ confidence levels with red dashed lines. KDE contour levels at 1, 2, and 3 $\sigma$-levels (increasing-inward) are represented with solid black lines. Blended lines are showed as light blue circles, multiple true lines predictions in black circles, and red circles mark cases where both occur. Other isolated detections of lines are denoted in light grey small circles. On the right, a normalized histogram of the count of occurrences of the different cases.}
    \label{fig:centers}
\end{figure}

\subsection{Reliability of the detection: position of the absorbers}\label{subsect:position}

As an additional test of the reliability of the detection, we measured the center of the detected lines. The line center is defined as the weighted mean wavelength over the detection range:
\begin{equation}
    \lambda_\mathrm{det} = \frac{\displaystyle\sum_{i = \mathrm{px}_i}^{\mathrm{px}_f} \lambda_i \left( 1 - f_i \right)}{\displaystyle\sum_{i = \mathrm{px}_i}^{\mathrm{px}_f} \left( 1 - f_i \right)}
\label{eq:lambda_det}
\end{equation}
where $f$ is the normalized flux, $\lambda$ the associated wavelength, and $\mathrm{px}_{i,f}$ are the initial and final bins of the detected lines.
Figure~\ref{fig:centers} shows the Bland–Altman plot showing the agreement between the true and predicted centers.
The average of the difference between true and predicted centers is $\left< \Delta \lambda_\mathrm{C} \right> \approx -0.028\,\mathrm{\AA}$, which corresponds to $25\%$ of a pixel size, with a standard deviation $\sigma_{\lambda_\mathrm{C}} \sim 0.205\,\mathrm{\AA}$.
The mean difference and its variations are minimal, indicating a good match between the true center of the line and the computed weighted center on the predicted line position.
$73.5\%$ of the predicted centers are recovered within $1$ pixel, and $49.7\%$ within $1/2$ of a pixel.
The Kernel Density Estimator (KDE) contour lines highlight the most dense regions near a zero mean difference, being mostly symmetrical, further demonstrating the reliability of the model's detections.
It is worth noting that some of the metal absorption lines are blended.
This can cause some of the dispersion seen, but overall it is a minor effect, since we compute the rate of blends over the entirety of lines to be $1.8\%$.
Even though blended lines are considered as separate systems in the metric computation, the predicted center is often associated with two different lines. Our analysis method for the position of the absorbers has the current limit of not being able to distinguish between lines in case of blended features.
In Fig.~\ref{fig:centers}, we can also notice the presence of multiple lines detections. These are the cases where, in the range of wavelengths over which the model predicted a line, there is more than one true absorption line.
In this case where there are present more true absorption lines in one single detection range, these lines are not necessarily blended. Likewise, the prediction range of a line that is blended often does not include the second blended line; therefore not falling into the category of multiple lines. However, the two cases can occur simultaneously, when in one single detection the line detected has a blend, and the blend is present under the predicted range.
Both blends and multiple lines findings are more common in the bluest side of the spectra, contributing to the dispersion mostly in that region.

\subsection{Detectability after continuum fitting}\label{subsec:cont_fit}

Up to now, we have focused on the performance of the network on idealised data for which the continuum level was perfectly known. As real applications also require a continuum-fitting step, we are interested next in understanding the U-Net performance on data that are continuum-normalized, starting from a complete set of mocks where the quasar continuum has been injected and then determined using the autoencoder presented by \citet{pistis2025continuum}.

\begin{figure}
    \resizebox{\hsize}{!}{\includegraphics[width=1.0\textwidth]{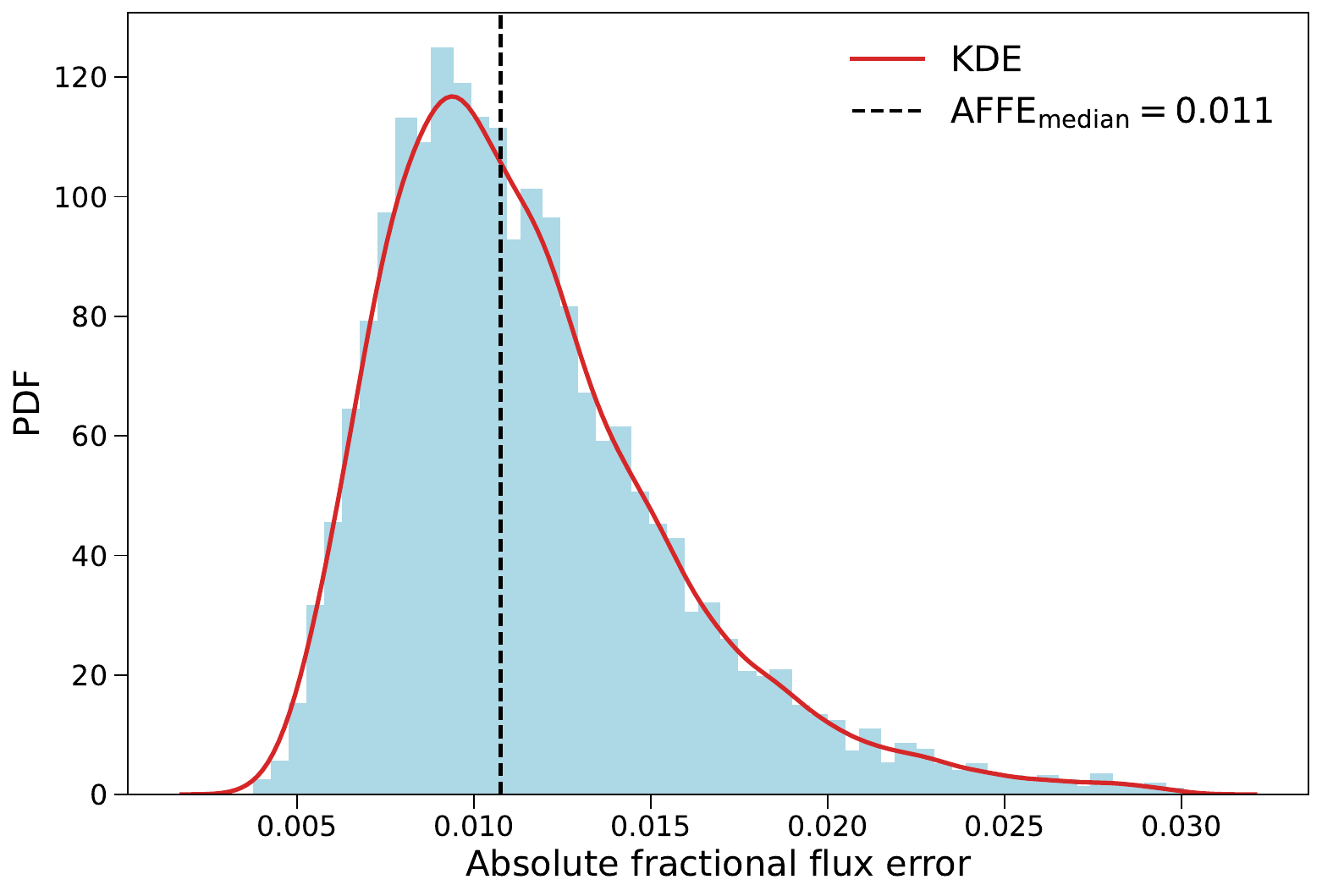}}
    \resizebox{\hsize}{!}{\includegraphics[width=1.0\textwidth]{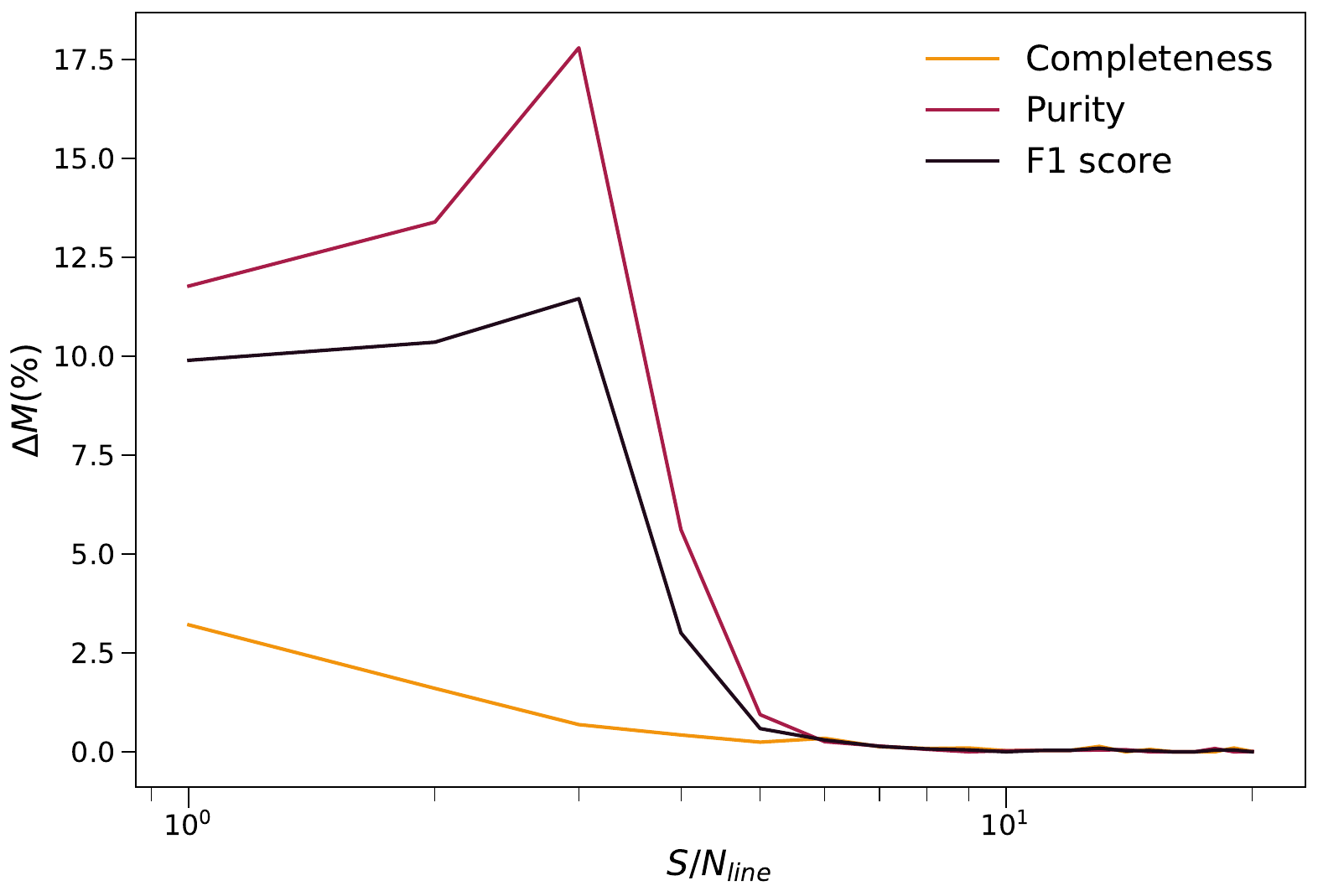}}
    \resizebox{\hsize}{!}{\includegraphics[width=1.0\textwidth]{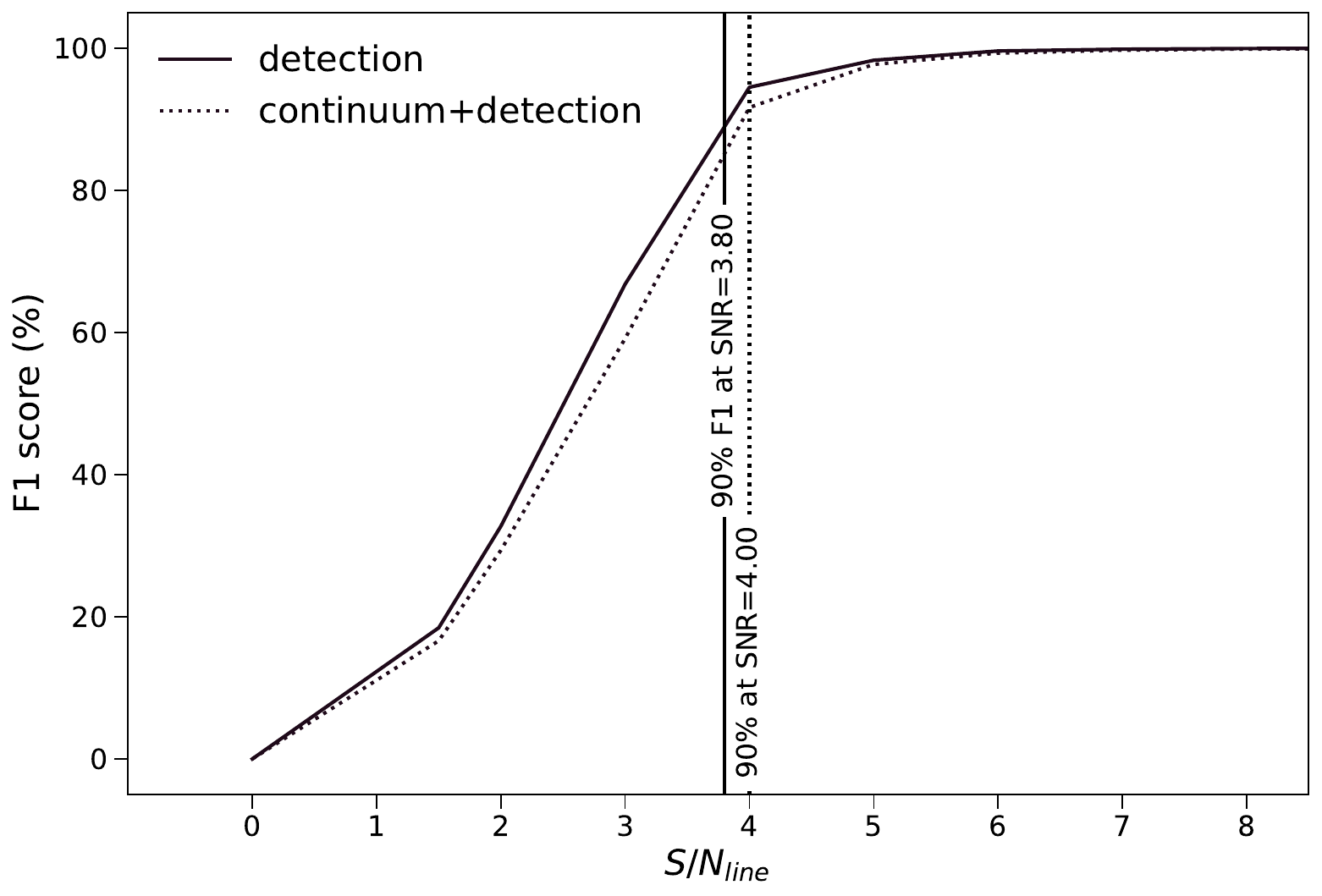}}
    \caption{Performance of the sequential use of the models for the continuum fitting and line detection. Top panel: histogram of the absolute fractional flux error of the autoencoder used for the continuum fitting. Middle panel: fractional difference of the various metrics (completeness, purity, F1 score) as a function of the signal-to-noise of the line ($\mathrm{S}/\mathrm{N}_\mathrm{line}$). Bottom panel: F1 score curves for the detection-only and continuum plus detection pipelines as a function of the line signal-to-noise ratio ($\mathrm{S}/\mathrm{N}_\mathrm{line}$). Values for $\mathrm{S}/\mathrm{N}_\mathrm{line}>8$ are omitted, as they are all approximately $100\%$.}
    \label{fig:cont_det}
\end{figure}

The autoencoder we use in this work is a slightly modified and retrained version of the one in \citet{pistis2025continuum}.
In Fig.~\ref{fig:cont_det} we reassess the performance of the autoencoder for the continuum fitting.
The model reproduces the true continuum with a median error of $\left< \mathrm{AFFE} \right> = 0.011$, computed over all the rest-frame wavelength range, that is in agreement with the value found in \cite{pistis2025continuum}. 
Having established that the retrained autoencoder yields comparable performances to the original one, we finally tested the full pipeline (continuum fitting plus line detection) needed for observational data.
To enable comparison with the results presented in earlier sections, we define the difference in a given metric (purity, completeness, and F1 score) as
\begin{equation}
    \Delta \mathrm{M} = \left| \frac{\mathrm{M}_\mathrm{d} - \mathrm{M}_\mathrm{c+d}}{\mathrm{M}_\mathrm{d}} \right|
\end{equation}
where $\mathrm{M}$ corresponds to the chosen metric while the suffixes $d$ and $c+d$ correspond to detection and continuum plus detection pipelines, respectively.
The horizontal axis, for the central and bottom panels, represents the line S/N under the ideal-continuum assumption, rather than a measurement significance incorporating continuum-fitting uncertainty. The small median AFFE is encouraging, but does not by itself establish that local continuum errors are negligible for all weak lines, exactly where the performances are deteriorated.

The middle panel of Fig.~\ref{fig:cont_det} shows the change of performance $\Delta \mathrm{M}$ when the step of the continuum fitting is added in the analysis. We note that the U-Net has not been retrained at this step, so the training is still performed on idealized data with no errors on the continuum level.
The bottom panel of Fig.~\ref{fig:cont_det} shows, for lower $\mathrm{S}/\mathrm{N}_\mathrm{line}$, the different F1 metrics for the detection only step, thus using the idealized continuum from the mocks, and for the continuum plus detection pipeline, using the estimated continuum from the autoencoder.
This comparison shows that using the estimation of the continuum from the autoencoder yields, as expected, slightly worse results than using mock spectra with a perfect normalization.
However, the difference in performance is limited to low $\mathrm{S}/\mathrm{N}_\mathrm{line}$ values, with $\Delta \mathrm{M}<10\%$ for the F1 score when $\mathrm{S}/\mathrm{N}_\mathrm{line}>3$. From $\mathrm{S}/\mathrm{N}_\mathrm{line}\approx 6$, the delta metric tends to $0$; at $\mathrm{S}/\mathrm{N}_\mathrm{line}\approx 4$, which we have established above as a reasonable threshold where the U-Net performs reliably, $\Delta \mathrm{M}\lesssim 5\%$ for the F1 score. The relative difference for lines of $\mathrm{S/N} \geq 4$ is reasonably small for purity ($\lesssim 10\%$) and negligible for completeness.
This test shows that using the continuum estimation from the autoencoder does not introduce significant deterioration in the model's ability to recover absorption lines. Given this, we can expect to apply the existing trained models of continuum fitting and detection of metal absorption lines to future WEAVE-QSO data consecutively.

\section{Discussion and comparison with other networks}\label{sec:comparison}

We presented a U-Net model able to successfully identify metal absorption lines in mock quasar spectra.
Our work adds a new architecture, the U-Net, to the growing body of literature that is employing machine learning for metal lines detection. For instance, CNNs have been used for similar tasks \citep[e.g., ][]{zhao_mgii, xia_caii, Szakacs_mgii, Liu_caii}.
Some of these works have explored the use of deep neural networks to find narrow absorption lines in quasar spectra, but have focused on a single metal species.
\citet{zhao_mgii} and \citet{Szakacs_mgii} focused on \ion{Mg}{ii} metal absorbers, while \citet{xia_caii} and \citet{Liu_caii} targeted \ion{Ca}{ii} absorbers.
A common trait in these works is the use of CNN as the architecture of choice, due to its widely established ability to detect features in images. A general conclusion from these applications is that using a first filter size comparable to the typical width of the absorption feature can improve the network performance.
This choice of the filter width implies that the networks should be tailored to the spectral resolution of the dataset when the line spread function dominates the profile (in moderate and lower resolution data). A question, however, arises on the performance of these models on much higher resolution data, when absorption lines will be dominated by a variety of line widths that are intrinsic and no longer driven by the resolution of the spectrograph.

In our case, since we explore multiple metal ion species with varying EW, the filter size was allowed to vary (see Sec.~\ref{subsec:abs_det}).
This does not pose a problem for the success of our results; in fact, the U-Net uses skip-connections to help preserve information about the context around the lines to be able to compensate for the loss of spatial information typical of CNNs.

\begin{table*}
\caption{Summary table listing pixel size,  $\mathrm{S}/ \mathrm{N}$ selection, metal ion targets, training set size and computed metrics (F1 score:F1, Completeness:C, Purity:P, Accuracy:a) of this work compared to the works discussed in Sect. \ref{sec:comparison}.}
\label{tab:lit_comparison}
\centering
\begin{tabular*}{\textwidth}{l @{\extracolsep{\fill}} l l l l l}
\hline\hline
\noalign{\smallskip}
Work & Pixel size &  $\mathrm{S}/ \mathrm{N}$ Selection & Metal ion targets & Training set size & Metric\\
\hline
\noalign{\smallskip}
This work & $\sim0.114\AA$ & $\mathrm{S}/ \mathrm{N}_\mathrm{line} \geq 1.5$ & (refer to Table \ref{tab:metal_ions}) & $35\,000$ & $F1\sim 82\%$\\
&&&&&$C=92\%$\\
&&&&&$P=74\%$\\
 & & $\mathrm{S}/ \mathrm{N}_\mathrm{line} \geq 4 $&  &  & $F1=C=P\sim99\%$\\
\hline
\noalign{\smallskip}
\citet{zhao_mgii} & $\sim0.4\AA$ & $EW_{2796/2803}\geq0.3\AA$  & \ion{Mg}{ii} & $34\,379^\dagger$ & $a\sim 94\%$\\
\citet{Szakacs_mgii} & $R=20\,000^*$ & $\mathrm{S}/ \mathrm{N}_\mathrm{spectrum} \geq 3$ & \ion{Mg}{ii} & $108\,000$ & $a\sim 98.6\%$\\
\citet{xia_caii} & $1\AA$ & $\mathrm{S}/ \mathrm{N}_\mathrm{\lambda3934} \geq 2.5 $ & \ion{Ca}{ii} & $680\,000$ & $a=F1\sim 95.9\%$\\
 &  & $\mathrm{S}/ \mathrm{N}_\mathrm{\lambda3969} \geq 2 $ &  &  & \\
\citet{Liu_caii} & $1\AA$ & $\mathrm{S}/ \mathrm{N}_\mathrm{\lambda3934} \geq 2.5$ & \ion{Ca}{ii} & $216\,000$ & $a=F1=C=P$\\
&&$\mathrm{S}/ \mathrm{N}_\mathrm{\lambda3969} \geq 2 $&&&$\sim99.77\%$\\
\hline
\end{tabular*}
\tablefoot{
\tablefoottext{*}{ \citet{Szakacs_mgii} relies on high-resolution mock spectra with the indicated resolution $R$.}
\tablefoottext{$\dagger$}{\citet{zhao_mgii} use $5$ train sets of the same size with the same spectra containing \ion{Mg}{ii}, but different control spectra. The indicated number is the total sum of the $5$ train sets.}
}
\end{table*}

The previously cited works obtained high performances, with an accuracy of $a\sim94\%$ for \citet{zhao_mgii}, an accuracy and F$1$ score of $\sim95.9\%$ for \citet{xia_caii}, an accuracy of $a\sim98.6\%$ for \citet{Szakacs_mgii}, and $a\sim99.77\%$ for \citet{Liu_caii}.
The accuracy is the ratio between the correct classifications over the total classifications.
These values are compatible with our results taking into consideration only lines for which the $\mathrm{S}/ \mathrm{N}_\mathrm{line}\geq 4$, obtaining total metrics of $F1=C=P\sim99\%$.

Overall, we obtain a F$1$ score of $F1=82\%$, completeness of $C=92\%$, and purity of $P=74\%$, slightly lower results compared to the previous works.
Our total metrics are derived from predictions over quasars spectra regardless of their $\mathrm{S}/ \mathrm{N}_\mathrm{spectrum}$, performing a selection only over the metal absorption lines signal-to-noise ratio ($\mathrm{S}/ \mathrm{N}_\mathrm{line} \geq 1.5$, Sect. \ref{subsec:preprocessing}). They are, therefore, dominated by the majority of low $\mathrm{S}/ \mathrm{N}_\mathrm{line}$ lines.

Nevertheless, previous works were also subject to selection. \citet{Szakacs_mgii} worked with spectra having a $\mathrm{S}/ \mathrm{N}_\mathrm{spectrum}$ taken from a discrete uniform distribution with a minimum $\mathrm{S}/ \mathrm{N}_\mathrm{spectrum}$ equal to $3$. Furthermore, \citet{zhao_mgii} filtered out \ion{Mg}{ii} absorption lines with $EW_{2796/2803} < 0.3 \mathrm{\AA}$, on the other hand \citet{xia_caii} and \citet{Liu_caii} instead inserted only \ion{Ca}{ii} lines with $\mathrm{S}/ \mathrm{N}_\mathrm{\lambda3934} \geq 2.5$ and $\mathrm{S}/ \mathrm{N}_\mathrm{\lambda3969} \geq 2$ in their artificial spectra dataset.

Given the substantial differences in spectral resolution, targeted ion species, training-set selection, and task generality, the comparison in Table~\ref{tab:lit_comparison} should be read as contextual rather than as a controlled ranking of methods. It should be noted that the metrics used in \cite{Szakacs_mgii, Liu_caii} measure spectrum-level binary classification, i.e.,  whether an absorber is present in a given spectrum; whereas our completeness, purity, and F1 are computed at the level of individual detected lines across the full spectral range and multiple ion species.
The fraction of correctly labeled pixels or events over all examples is conceptually distinct from the line-level completeness, purity, and F1 score we report. In particular, accuracy is sensitive to class imbalance (most pixels are non-absorbers) and can appear artificially high even when rare absorption features are poorly recovered. Completeness and purity are instead computed at the level of individual detected lines and directly quantify the network's ability to find real absorbers and suppress false positives, respectively.

Concerning the wavelength accuracy, we obtain a mean absolute error $MAE\simeq0.11\,\mathrm{\AA}$ for a pixel size of $\sim 0.114\ \mathrm{\AA}$, measuring the line center as defined in Equation~\ref{eq:lambda_det}, whereas \citet{Szakacs_mgii} obtained an overall accuracy of $MAE=6.9\,\mathrm{\AA}$ with a resolution of the spectra of $R=\frac{\lambda}{\Delta\lambda}=20000$.

A comparison of performances also needs to take into account the degree of specialization of the various networks. 
\citet{Szakacs_mgii} considered only synthetic spectra containing only one \ion{Mg}{ii} doublet within each spectrum, excluding the presence of both other \ion{Mg}{ii} absorbers in the spectra and other metals' absorption lines. Applicability to more general cases is likely to require new training or changes to the architecture.
The models by \citet{xia_caii} and \citet{Liu_caii} require instead knowledge of the positions of the \ion{Mg}{ii} absorption lines to be able to detect \ion{Ca}{ii} lines. Lastly, \citet{zhao_mgii} searched for \ion{Mg}{ii} features only in the spectral region between the \ion{C}{iv} and \ion{Mg}{ii} emission lines.

As illustrated in Sect. \ref{subsec:abs_det}, we preferred the U-Net model in comparison to attention based architectures. It is important to consider the fact that \citet{Liu_caii}'s final choice of architecture involves the introduction of an attention mechanism at the end of each residual block's final convolution unit. They argue that this choice improves their detection abilities compared to previous work also on new unseen data, obtaining a  precision of $92.6\%$ compared to that of $20.3\%$ of \citet{xia_caii}.
Given this result, the introduction of an attention mechanism in a future updated version of this work could be beneficial.

Compared to these previous works, our U-Net aims to generalize the problem to the detection of any metal absorption lines (see Table~\ref{tab:metal_ions} for the list of ions included in our mocks) and considers the full spectral range redward of the Ly$\alpha$ emission line. This is a more general and hence challenging task; thus, the U-Net performance appears satisfactory even if purity and completeness start to drop below $90\%$ for $\mathrm{S}/\mathrm{N}_\mathrm{line}\lesssim 4$. Unlike other works, however, our architecture is specifically designed to identify absorption features without recognizing and classifying the specific ions. 
The U-Net model we have developed is intended to form part of a pipeline for the identification of metal ions absorption lines in quasar spectra.
The pipeline comprises three distinct steps, which are delineated in \citet{pistis2025continuum}, this paper, and \citet{pistis2026pilot}, respectively.
\citet{pistis2025continuum}, as mentioned throughout the manuscript, describes the continuum fitting step procedure for the fitting of the quasar continuum with an autoencoder architecture.
This procedure constitutes a significant pre-processing step, which has also been applied in this study (see Sect. \ref{subsec:cont_fit}).
The work presented here serves as the next pre-processing step: using the continuum given as the output of the autoencoder, the spectra are normalized and used as the input of the U-Net model, which in return flags the pixels corresponding to metal absorption lines.
Lastly, the original flux, the autoencoder continuum, and the pixel-level detection map are entered into a classification algorithm, as delineated in \citet{pistis2026pilot}, which performs an automatic classification of metal absorption lines.

\citet{pistis2026pilot} also provide a validation of the pipeline using real observational data, specifically from the first public data release of DESI \citep[DESI DR$1$][]{desi2025dr1}. The characteristics of DESI survey differ from the WEAVE-like mocks considered here in terms of resolution ($\rm{R_{DESI}}=2000-5500$), noise properties, and sample size. While our model is evaluated under idealized conditions, the application to DESI data incorporates observational effects such as bad pixels and sky residuals.
Another aspect to consider is the presence of broad absorption line (BAL) quasars, both in future WEAVE-QSO and DR$1$ spectra. According to \citet{2022MNRAS.511.3514E}, $\sim 12-16\%$ of quasar spectra in large survey present BAL features, which can be a potential source of contamination.
\citet{pistis2026pilot} tackles this problem by using a BAL finding tool to select BAL-free spectra before applying the full pipeline.
The validation presented in the work of \citet{pistis2026pilot} demonstrates that the pipeline maintains good performance when applied to real spectra, having an average $1.2\sigma$ tension with the literature catalogs for \ion{C}{iv} detections \citep{Cooksey_2013, hasan2020civ}, while also highlighting the additional uncertainties associated with real data. This population-level consistency is encouraging. However, the WEAVE-like mocks lack metal lines associated with systems with $14 \leq \log (N_{\ion{H}{i}}/{\mathrm{cm}^{-2}})\leq 16$, which can produce a non-detection bias for such systems.

\section{Conclusions}\label{sec:conclusions}

This study focuses on building an NN model for the detection of absorption lines in quasar spectra arising from metals in the IGM or CGM along the line of sight.
We select the U-Net as our architecture for its ability to classify small-scale features in the context of larger-scale properties of the dataset. This capability appears well-suited to the task of identifying narrow absorption lines within the quasar spectrum at greater wavelengths than the Ly$\alpha$ emission line, where metal lines are present.

We train the model on realistic mock spectra that mimic the data products of the WEAVE surveys and include a wide variety of ions, which have real-looking absorption profiles and are distributed in wavelengths according to observed redshift distributions.
It is possible in future to introduce further realism to the mocks, expanding the metal absorption lines systems injected into the spectra.
After training on idealised spectra that do not include the quasar continuum (i.e., corresponding to a perfect continuum level), the model reaches high levels of completeness and purity, above $90\%$, at $\mathrm{S}/\mathrm{N}_\mathrm{line} \approx 4$.
At high signal-to-noise, the U-Net is fully complete and suffers only from the detection of occasional false positives that lie in the wings of strong absorbers, such as DLAs (see Fig.~\ref{fig:spectra_examples}).

The model is also able to accurately center the absorption lines. The simple calculation of the centers from the weighted (using the absorption level as weight) mean of the wavelengths within a line find an average difference between true and predicted centers $\left< \Delta \lambda_\mathrm{C} \right> = -0.028$ with a standard deviation $\sigma_{\lambda_\mathrm{C}} = 0.205$ (see Fig.~\ref{fig:centers}).

We also explore, without retraining, the U-Net performance to applications in which the quasar continuum is measured on the data, and hence not fully idealized. This is achieved by applying the quasar continuum to our mocks, and renormalizing the spectra with the autoencoder by \citet{pistis2025continuum}, which we specialize to work across the entire wavelength range redward of the quasar Ly$\alpha$ line.

Even with this dataset, the performance of the network remains satisfactory. Compared to the more idealized case of a perfect continuum, the  U-Net shows no appreciable difference in completeness, a purity that is within $\approx 10\%$ of the idealized case, and a F$1$ score within $\approx 5\%$ for $\mathrm{S}/\mathrm{N}_\mathrm{line} \approx 4$. By $\mathrm{S}/\mathrm{N}_\mathrm{line} \approx 6$, the performance with and without continuum normalization becomes identical (see Fig.~\ref{fig:cont_det}).
Our models can thus detect lines with good precision of their centers while maintaining high performance despite uncertainty in the continuum fit.

This work expands on the set of tools available for the identification of absorption lines in large spectroscopic surveys, extending previous models in the capability to identify any absorption feature. Our U-Net is, however, not trained to distinguish and classify different ions, a task we will explore in a companion paper \citep{pistis2026pilot}. 

\section*{Data Availability}
Supplementary materials with the trained models for all architectures are available on Zenodo for both quasar continuum fitting and metal absorption line detection at \hyperlink{https://doi.org/10.5281/zenodo.22711575}{https://doi.org/10.5281/zenodo.22711575}.

\begin{acknowledgements}
We thank the referee for insightful comments that have improved the content and presentation of this work. 
This work has been supported by the European Union -- Next Generation EU, Mission 4, Component 1 CUP H53D23011030001.
IPR was supported by funding from the grant PID2023-151122NA-I00 by MICIU/AEI/10.13039/501100011033 and by ERDF/EU.
\end{acknowledgements}

\bibliographystyle{aa}
\bibliography{references}

\onecolumn
\begin{appendix}

\section{Training data and model architecture selection}\label{app:selection}
\begin{figure}[ht]
    \centering
    \includegraphics[width=0.82\textwidth]{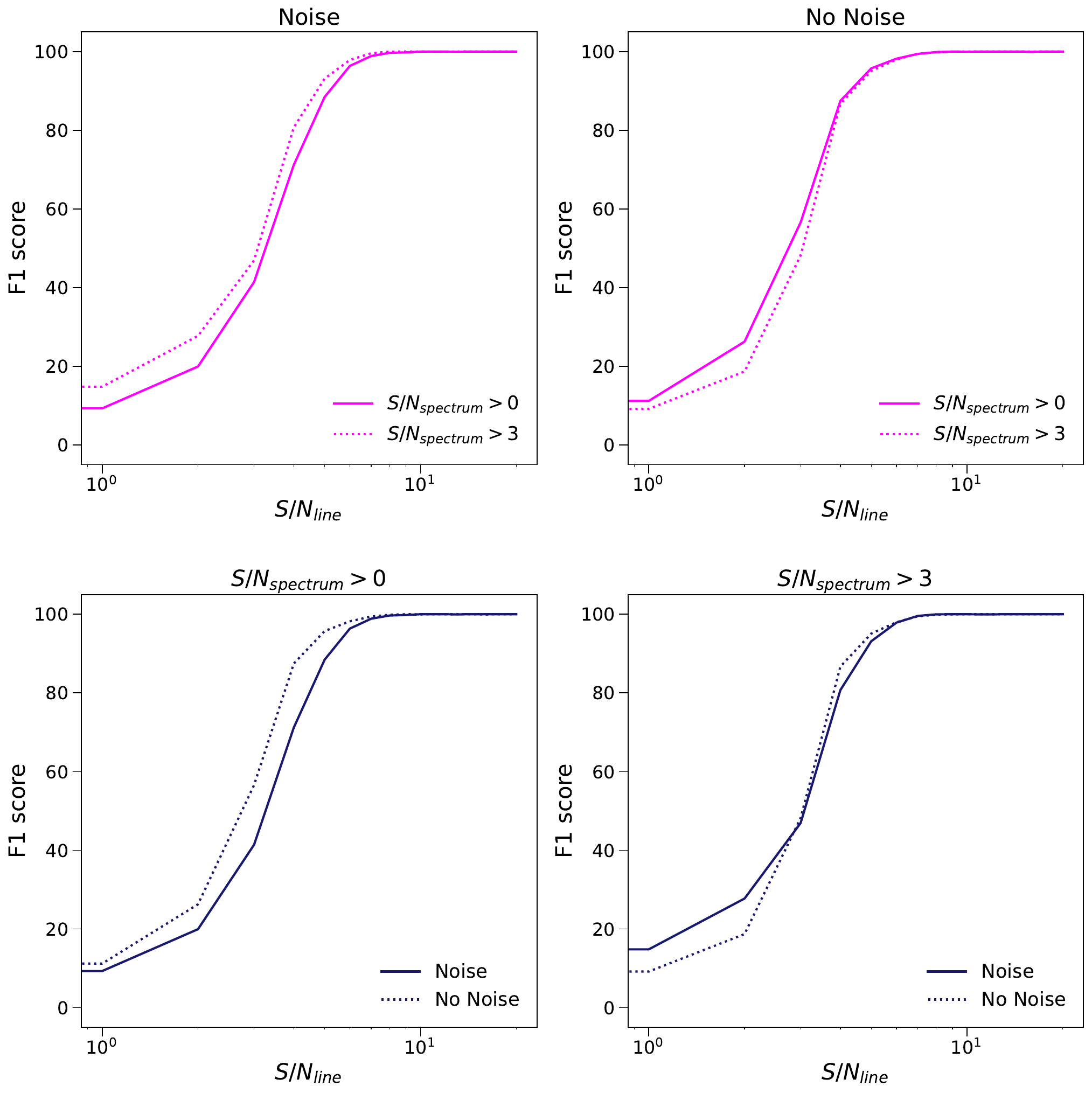}
    \caption{F1 score metric trends in function of the $\mathrm{S}/ \mathrm{N}_\mathrm{line}$ of the metal absorption lines. Top plots: F1 score of the sample with $\mathrm{S}/ \mathrm{N}_\mathrm{spectrum}>3$ spectra (dotted line) and with all spectra (solid line) using the model with (left panel) and without noise (right panel). Bottom plots: F1 score using the model without (dotted line) and with noise (solid line) on the data with all the spectra (left panel) and with only the high $\mathrm{S}/ \mathrm{N}_\mathrm{spectrum}>3$ spectra (right panel).}
    \label{fig:selection}
\end{figure}

To train the U-Net model discussed in Sect.~\ref{subsec:abs_det}, we utilized all generated mock spectra instead of only those with $\mathrm{S}/ \mathrm{N}_\mathrm{spectrum}>3$, which would feature clearer absorption lines. We also tested a modified U-Net architecture to include noise and flux as inputs to enhance detection. 
Figure~\ref{fig:selection} shows F1 score metrics for the different training data with both U-Net models. The model with noise performs better with high $\mathrm{S}/ \mathrm{N}_\mathrm{spectrum}$ spectra, whereas the noise-free model shows improved results with all available spectra.
When using the noise model, the noise input channel introduces confusion when the $\mathrm{S}/ \mathrm{N}_\mathrm{spectrum}$ is low, resulting in a poorer performance compared to its use on higher $\mathrm{S}/ \mathrm{N}_\mathrm{spectrum}$ spectra.
In the comparison, the model without noise is preferred overall, particularly when using all spectra. For $\mathrm{S}/ \mathrm{N}_\mathrm{spectrum}>3$, the noise model is favorable only up to $\mathrm{S}/ \mathrm{N}_\mathrm{spectrum}=3$; beyond that, the noise-free model outperforms it.
Since we consider absorption lines with $\mathrm{S}/ \mathrm{N}_\mathrm{line} \geq 1.5$, the model without noise is considered the preferred model.
\noindent
We also previously consider a series of U-Net models, with architectures similar to the one we take into consideration, trained on sub-sets of data.
Spectra were divided into discrete ranges of magnitude and, following the same idea of considering part of the continuum absorption lines lower than the $\mathrm{S}/ \mathrm{N}_\mathrm{line}$ threshold, into bins of thresholds of $\mathrm{S}/ \mathrm{N}_\mathrm{line}$ of the absorption lines.
Training in different magnitude conditions and considering only some absorption lines makes it difficult to decide which model can generalize better to all different $\mathrm{S}/ \mathrm{N}_\mathrm{line}$ lines.
We also tested if the choice of shifting all the spectra into a common reference frame of the quasar rest wavelength range has an advantage over using the observed wavelength range.
In the rest frame there are fixed ranges of wavelengths in which the absorption lines from a given metal ion appear, always at a shorter wavelength than the corresponding rest frame emission peak. In the observed frame, instead, these regions are at different wavelengths.
We, indeed, observed that the common rest frame yields better results than using the different observed frames.     
\end{appendix}

\end{document}